\documentclass[aps,prm,prbib,twocolumn,superscriptaddress]{revtex4-2}
\usepackage{graphicx}
\usepackage{bm}
\usepackage{hyperref}
\usepackage{todonotes}
\usepackage{nicefrac}
\usepackage{bbm}
\usepackage[normalem]{ulem}
\usepackage{booktabs}
\usepackage{subfigure}
\usepackage{rotating}
\usepackage{color}
\usepackage{float}
\usepackage{amsmath}
\usepackage{multirow} 
\usepackage[switch,columnwise]{lineno}
\usepackage{tabularx}

\begin{document}

\title{Realistic tight-binding model for V$_2$Se$_2$O-family altermagnets}

\author{Xingkai Cheng}
\affiliation{Department of Physics, The Hong Kong University of Science and Technology, Clear Water Bay, Kowloon, Hong Kong SAR}

\author{Yifan Gao}
\affiliation{Department of Physics, The Hong Kong University of Science and Technology, Clear Water Bay, Kowloon, Hong Kong SAR}

\author{Junjie Peng}
\affiliation{Department of Physics, The Hong Kong University of Science and Technology, Clear Water Bay, Kowloon, Hong Kong SAR}

\author{Chunyu Wan}
\affiliation{Department of Physics, The Hong Kong University of Science and Technology, Clear Water Bay, Kowloon, Hong Kong SAR}

\author{Junwei Liu}
\email[]{liuj@ust.hk}
\affiliation{Department of Physics, The Hong Kong University of Science and Technology, Clear Water Bay, Kowloon, Hong Kong SAR}
\affiliation{IAS Center for Quantum Matter, The Hong Kong University of Science and Technology, Hong Kong SAR, China}

\date{\today}

\begin{abstract}
Following earlier theoretical prediction, intercalated V$_2$Se$_2$O-family altermagnets such as RbV$_2$Te$_2$O and KV$_2$Se$_2$O have now been experimentally confirmed as $d$‑wave altermagnets, representing the only known van der Waals layered altermagnetic systems. By combining crystal-symmetry-paired spin‑momentum locking (CSML) with the layered structure,
the V$_2$Se$_2$O-family offers suitable platform for studying low‑dimensional spintronic responses and exploring the interplay among multiple quantum degrees of freedom.
To establish a concrete theoretical foundation for understanding and utilizing these materials, we investigate six representative members of the V$_2$Se$_2$O-family and construct a realistic tight‑binding model parameterized by first‑principles calculations, which is benchmarked by experimental measurements. This model accurately captures essential altermagnetic electronic properties, including CSML and noncollinear spin-conserved currents. It further incorporates strain‑coupling parameters, enabling the simulation of strain‑tunable responses such as the piezo‑Hall effects. This realistic model allows systematic exploration of multiple degrees of freedom (like spin, valley, and layer) within a single system, and lays the groundwork for understanding their coupling with other quantum materials, such as topological insulators and superconductors, thereby advancing both the fundamental understanding and potential device applications of this novel class of layered altermagnets.
\end{abstract}

\maketitle

\section{Introduction}
Altermagnets (AMs) uniquely combine nonrelativistic spin-splitting with zero net-magnetization, thereby merging the advantages of both ferromagnets and antiferromagnets \cite{wuFermiLiquidInstabilities2007a,maMultifunctional2021a,tang_splitting_2014,hayami_splitting_2019,yuanGiant2020a,yuanPrediction2021a,smejkalConventional2022a,hu_2025,smejkalEmerging2022a,cheongAM2024,cheongAM2025}. These novel properties originate from specific crystal symmetry $\mathcal{C}$ (rotation or mirror), which simultaneously connects antiparallel magnetic sublattices in real space and enforces $\mathcal{C}$-paired spin–momentum locking (CSML) in reciprocal space \cite{maMultifunctional2021a}. This fundamental mechanism gives rise to various novel phenomena, including noncollinear spin-conserved currents \cite{maMultifunctional2021a,wuFermiLiquidInstabilities2007a,KB_SST_2025,naka_spin_2019,gonzalez-hernandezEfficientElectricalSpin2021a}, unconventional itinerant piezomagnetism \cite{maMultifunctional2021a,hu_2025,KB_strain2025}, magneto-optical effects \cite{zhouMO2021} and giant/tunnel magnetoresistance \cite{shaoSpinneutral2021a,smejkalGiant2022a}. These $\mathcal{C}$-governed features in both real and momentum space establish AM as a highly promising platform for generating, manipulating, and detecting spin responses, holding significant potential for next-generation spintronic devices \cite{liu_review_2024,song_review_2025}.

However, experimental realization of altermagnetic candidates, particularly those capable of efficiently generating spin currents, remains severely limited. This is primarily due to stringent symmetry and magnetic-order requirements. Specifically, the generation of nonrelativistic spin current necessitates anisotropic spin-polarized conductivity, which can be eliminated by rotational symmetry higher than two-fold within the same magnetic sublattice \cite{maMultifunctional2021a,gonzalez-hernandezEfficientElectricalSpin2021a}. Consequently, $g$-wave altermagnets like MnTe \cite{MnTe_hajlaoui_2024,MnTe_krempasky_2024,MnTe_lee_2024,KB_MnTe_2025} and CrSb \cite{CrSb_reimersDirect2024,CrSb_zengObservation2024a,zhou_crsb_2025,CrSb_lu_2025,CrSb_yangThreedimensional2025a}, exhibit isotropic conductivity due to their three-fold rotation, thus are excluded as effective spin sources. Additional complications arise from intricate magnetic orders: for instance, the magnetic ground state of RuO$_2$ is still debated \cite{ruo2_2017,ruo2_2019,ruo2_2024}, while MnTe$_2$ exhibits noncoplanar magnetic order breaking spin conservation \cite{MnTe22024}. Moreover, most reported altermagnets are three-dimensional bulk crystals that cannot be easily exfoliated into two-dimensional forms, limiting their integration into nanoscale devices and their potential coupling with other quantum materials such as superconductors and topological insulators \cite{SC_wangCoexistence2012}. In this context, the recent discovery of the room-temperature layered $d$-wave altermagnet V$_2$Se$_2$O-family, including Rb-intercalated V$_2$Te$_2$O and K-intercalated V$_2$Se$_2$O, presents a highly desirable system. 

In 2021, layered V$_2$X$_2$O (X = Te, Se) compounds were theoretically identified as altermagnets capable of noncollinear spin currents and unconventional piezomagnetism originating from CSML, even without spin‑orbit coupling (SOC) \cite{maMultifunctional2021a}. Recently, alternating spin polarization between symmetry‑paired valleys, a hallmark of CSML, have been observed in both Rb-intercalated V$_2$Te$_2$O \cite{rvto_2025,RVTO_2026} and K-intercalated V$_2$Se$_2$O \cite{kvso_2025} via spin-resolved angle-resolved photoemission spectroscopy. The altermagnetic order in these materials has been further confirmed by atomic‑scale scanning tunneling microscopy in both real and momentum space \cite{KVSO_STM_2025,CVSO_STM_2025}. Moreover, the observed negligible $k_z$ dispersion indicates the two‑dimensional character \cite{RVTO_2026}. The identified altermagnetic order, combined with the unique noncollinear spin-conserved currents and the van der Waals layered structure, offers a unique and compelling platform for both fundamental investigation and practical applications in low-dimensional spintronics. Therefore, a realistic model is essential to systematically understand their intrinsic properties, practical potential, and coupling effects with other quantum materials, such as topological superconductors realized via the superconducting proximity effect \cite{SC_wangCoexistence2012} and moiré superlattices \cite{caoUnconventional2018}. 

In this work, we develop a realistic tight‑binding (TB) model for six representative members of the V$_2$Se$_2$O-family, with parameters fitted from first‑principles calculations, ensuring consistency with experimental data \cite{rvto_2025,kvso_2025}. This model accurately reproduces the CSML‑governed electronic properties of V$_2$Se$_2$O-family in both absence and presence of SOC, and enables the simulation of novel responses such as noncollinear spin currents, layer polarization and piezo-Hall effect \cite{maMultifunctional2021a,hu_2025,zhou_crsb_2025,han_mnsi_2024}. Moving beyond minimal toy models, our realistic TB model serves as a versatile and predictive computational platform for investigating coupling effects among multiple degrees of freedom (including spin, valley, and layer), thus advancing both fundamental understanding and device applications in this emerging class of layered altermagnet family.

The remainder of the paper is organized as follows. Section \ref{tb-model} discusses the construction of the TB model and calculated CSML band structures with and without SOC. Section \ref{efield} presents the response to electric fields, including the emergence of noncollinear spin currents and layer polarization under an out‑of‑plane field. Section \ref{strains} introduces a strain‑coupling term to the model and demonstrates the resulting piezo‑Hall response. Finally, Section \ref{summary} provides a summary and outlines future research directions.

\section{Four-band TB model}\label{tb-model}

\subsection{TB model without SOC}

\begin{figure}[tbp!]
	\centering
	\includegraphics[width=0.9\linewidth]{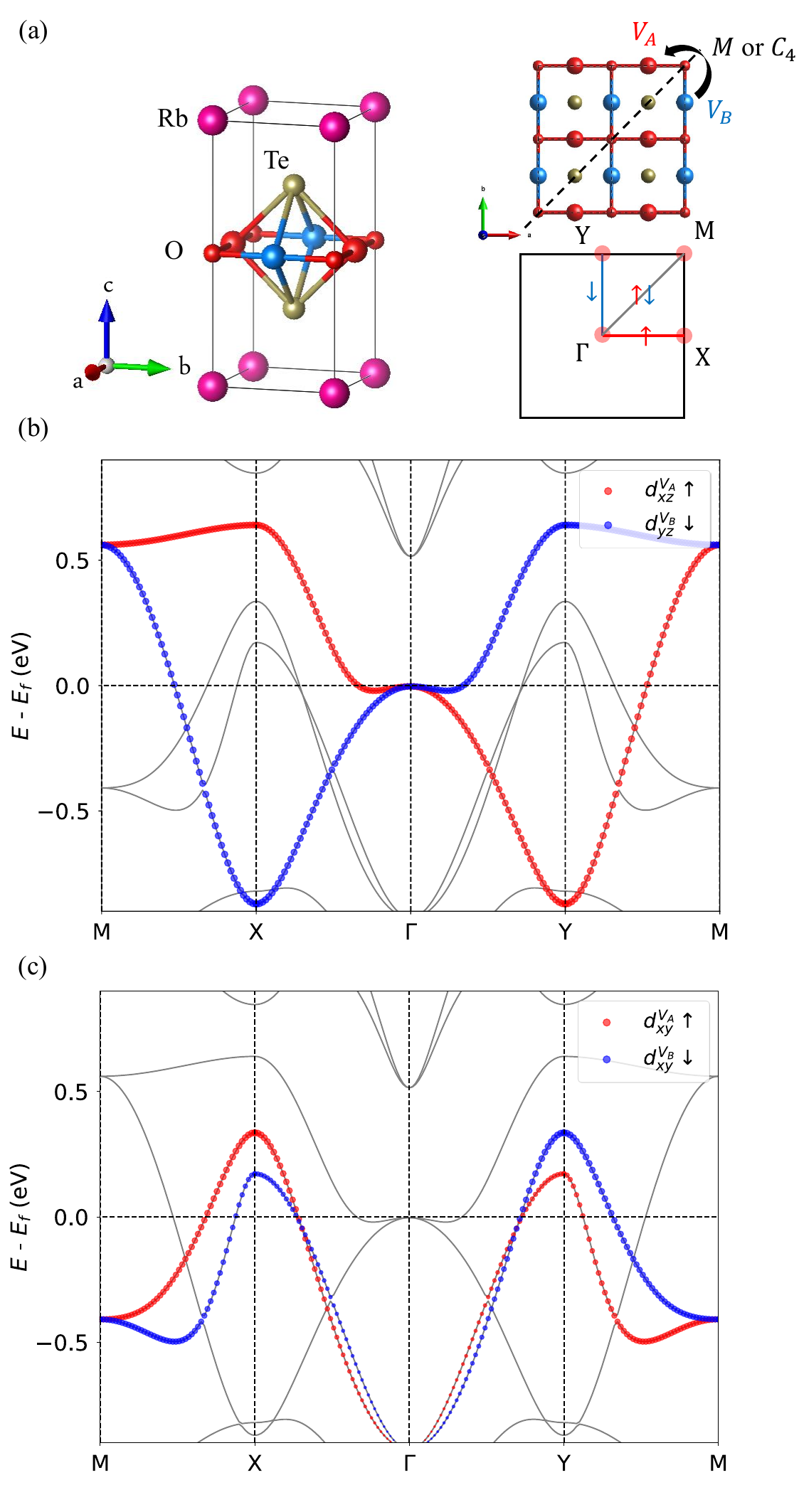}
	\caption{{\bf Lattice and CSML for Rb-intercalated V$_2$Te$_2$O.} (a)  Crystal structure and Brillouin zone of Rb-intercalated V$_2$Te$_2$O, where the two sublattices are connected by $\mathcal{C}_4$ and mirror symmetries. (b)-(c) Orbital‑projected CSML band structures for (b) $d_{xz}$ orbitals with spin-up from V$_A$ and $d_{yz}$ orbitals with spin-down from V$_B$ and (c) $d_{xy}$ orbitals with spin-up from V$_A$ and $d_{xy}$ orbitals with spin-down from V$_B$. Red and blue circles refer to spin-up and spin-down, respectively.}
	\label{fig:lattice}
\end{figure}

The space group of V$_2$X$_2$O is P4/mmm, which contains two vanadium atoms (denoted as V$_A$ and V$_B$) per unit cell \cite{maMultifunctional2021a,rvto_2025,rvto_2018}. As shown in  Fig.~\ref{fig:lattice}(a), the two V sublattices are connected by a four-fold rotational symmetry $\mathcal{C}_{4z}$ and the mirror symmetries $\mathcal{M}_{xy}$ and $\mathcal{M}_{\bar{x}y}$ in real space, which simultaneously enforces a characteristic CSML in reciprocal space ( Fig.~\ref{fig:lattice}(b)). Moreover, the point group mmm of each sublattice allows for anisotropic spin‑polarized conductivity, thereby leading to noncollinear spin-conserved currents \cite{maMultifunctional2021a,rvto_2025}, which will be discussed in more detail in Sec.~\ref{efield}. Given the layered nature of V$_2$X$_2$O and experimentally confirmed negligible dispersion along the $k_z$ direction, our TB model focuses on the intralayer hopping. 

To determine appropriate basis for the TB model, we perform orbital projections for the bands near the Fermi level. Take the Rb-intercalated V$_2$Te$_2$O as an example,  Fig.~\ref{fig:lattice}(b) and (c) show that bands around the Fermi level are primarily contributed by $d_{xz}^{\uparrow}$ and $d_{xy}^{\uparrow}$ from the V$_A$, together with $d_{yz}^{\downarrow}$ and $d_{xy}^{\downarrow}$ from the V$_B$ (where $\uparrow$ and $\downarrow$ denote spin-up and spin-down, respectively), thus forming a special crystal symmetry-paired orbital orders in real space. Similar orbital compositions are observed in other V$_2$X$_2$O compounds, as detailed in Appendix~\ref{appendix_projection}. Therefore, we select these four dominant orbitals as the basis for our TB model, denoted by \{$d_{xz}^{A, \uparrow}, d_{xy}^{A,\uparrow}, d_{yz}^{B, \downarrow}, d_{xy}^{B, \downarrow}$\}. 

\begin{figure*}[t]
	\includegraphics[width=0.9\textwidth]{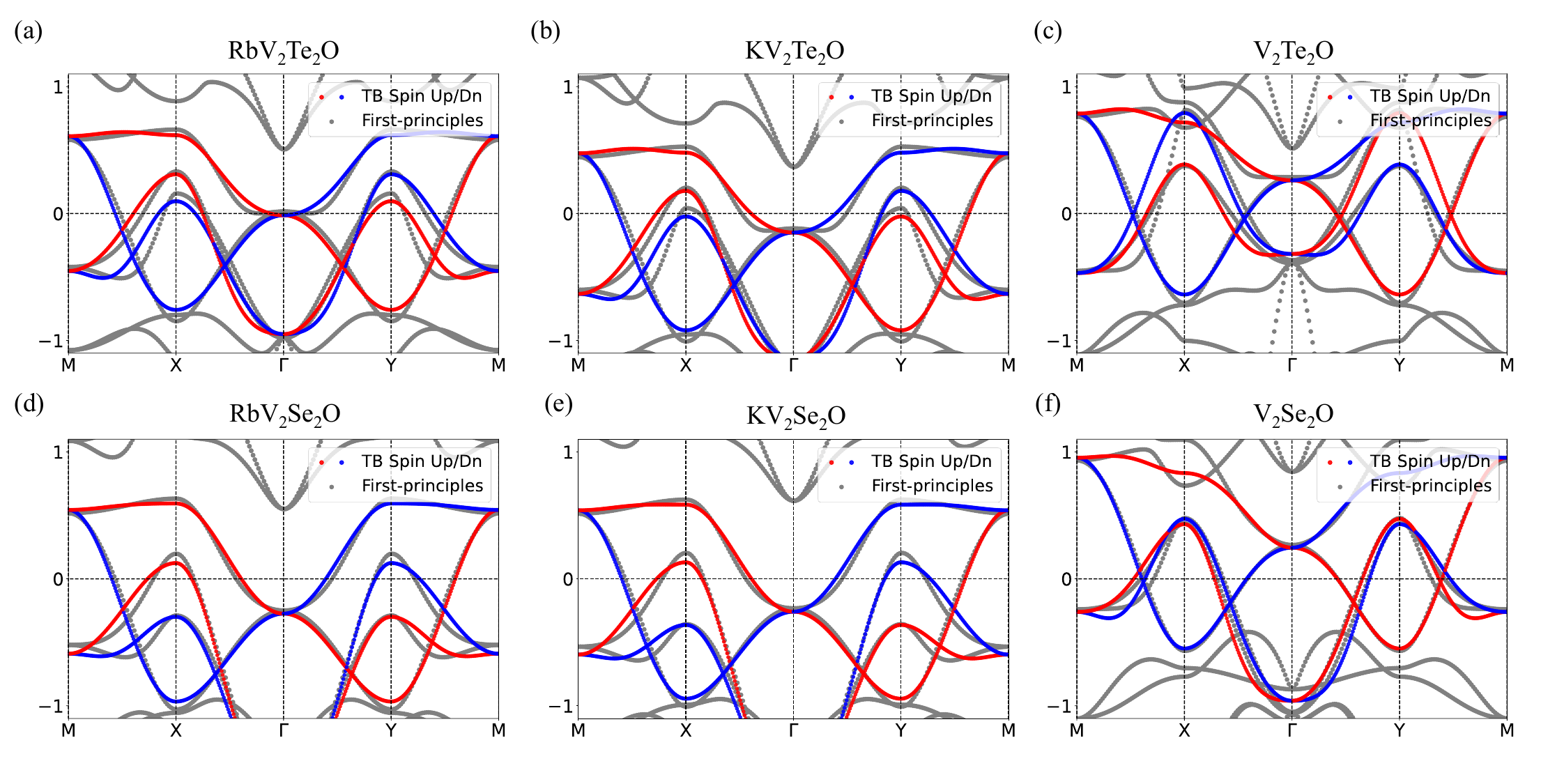}
	\caption{{\bf CSML band structures in the absence of SOC.} Red and blue dots denote the spin-up and spin-down band structures calculated by the TB model, respectively. Results from first-principles calculations are shown as grey dots for comparison.}
	\label{fig:nsoc}
\end{figure*}

Having determined the basis, the hopping integrals are defined as $t_{ij} = \langle \phi(\mathbf{r}_i)| \hat{H} |\phi(\mathbf{r}_j)\rangle$, where $\phi(\mathbf{r}_i)$ and $\phi(\mathbf{r}_j)$ represent the orbital basis states at the $i^{th}$ and $j^{th}$ lattice sites, respectively. Crucially, the crystal symmetry that connects the two magnetic sublattices enforces a $\mathcal{C}$-paired equivalence among the real-space hopping. For instance, the symmetry operation $\mathcal{M}_{\bar{x}y}$ exchanges the A and B sublattices in real space, while simultaneously transforming orbitals $d_{xz}^{A,\uparrow} \rightarrow d_{yz}^{B,\downarrow}$, thus imposing the hopping equivalence:  $  \langle d_{xz}^{A,\uparrow}(\mathbf{r}) | \hat{H} | d_{xz}^{A,\uparrow}(\mathbf{r} + a\hat{x}) \rangle = \langle d_{yz}^{B,\downarrow}(\mathbf{r}) | \hat{H} | d_{yz}^{B,\downarrow}(\mathbf{r} + b\hat{y}) \rangle$. This establishes a symmetry-dictated equivalence between the $x$-direction hopping on sublattice A and $y$-direction hopping on sublattice B.

All crystal symmetry-related hopping equivalences are summarized in Fig.~\ref{fig:hopping}, where we include hopping terms up to the third-nearest neighbors to balance simplicity and accuracy. The symmetry-restricted four-band TB Hamiltonian (without SOC) is expressed as:

\begin{equation}
H_0 = \begin{pmatrix}
H_{\uparrow} & 0 \\
0 & H_{\downarrow}
\end{pmatrix},
\end{equation}
where the spin sectors satisfy:
\begin{align}
H_{\uparrow} &= \begin{pmatrix}
H_{xz,\uparrow} & 0 \\ 
0 & H_{xy,\uparrow}
\end{pmatrix},\\
H_{\downarrow} &= \mathcal{M}_{\bar{x}y}^{-1} H_{\uparrow} \mathcal{M}_{\bar{x}y},
\end{align}
with \(\mathcal{M}_{\bar{x}y}\) being the crystal symmetry operation that connects the two opposite sublattices. The explicit matrix elements of the Hamiltonian are provided in Appendix~\ref{appendix_hopping}.

Here, we emphasize that crystal symmetry enforces a strictly diagonal structure of the tight-binding model, arising from two key mechanisms: (i) spin-up and spin-down blocks are decoupled in the absence of SOC, eliminating any off-diagonal terms between different spin blocks. (ii) within each spin block, the selected basis $d_{xy}^{A,\uparrow}$ ($d_{xy}^{B,\downarrow}$) and $d_{xz}^{A,\uparrow}$ ($d_{yz}^{B,\downarrow}$) exhibit opposite eigenvalues (+1 and -1, respectively) under $\mathcal{M}_z$ mirror symmetry. Those opposite eigenvalues prohibit both inter-orbital coupling and hybridization between different basis. This symmetry protection directly gives rise to the characteristic $C$-paired nodal lines (points) in altermagnets \cite{RVTO_2026}. We note that off-diagonal terms could emerge if additional orbitals or interlayer hopping along the \(c\)-axis were included, we defer such extensions to future studies due to the weak interlayer coupling.

\begin{figure*}[t!]
	\includegraphics[width=0.9\textwidth]{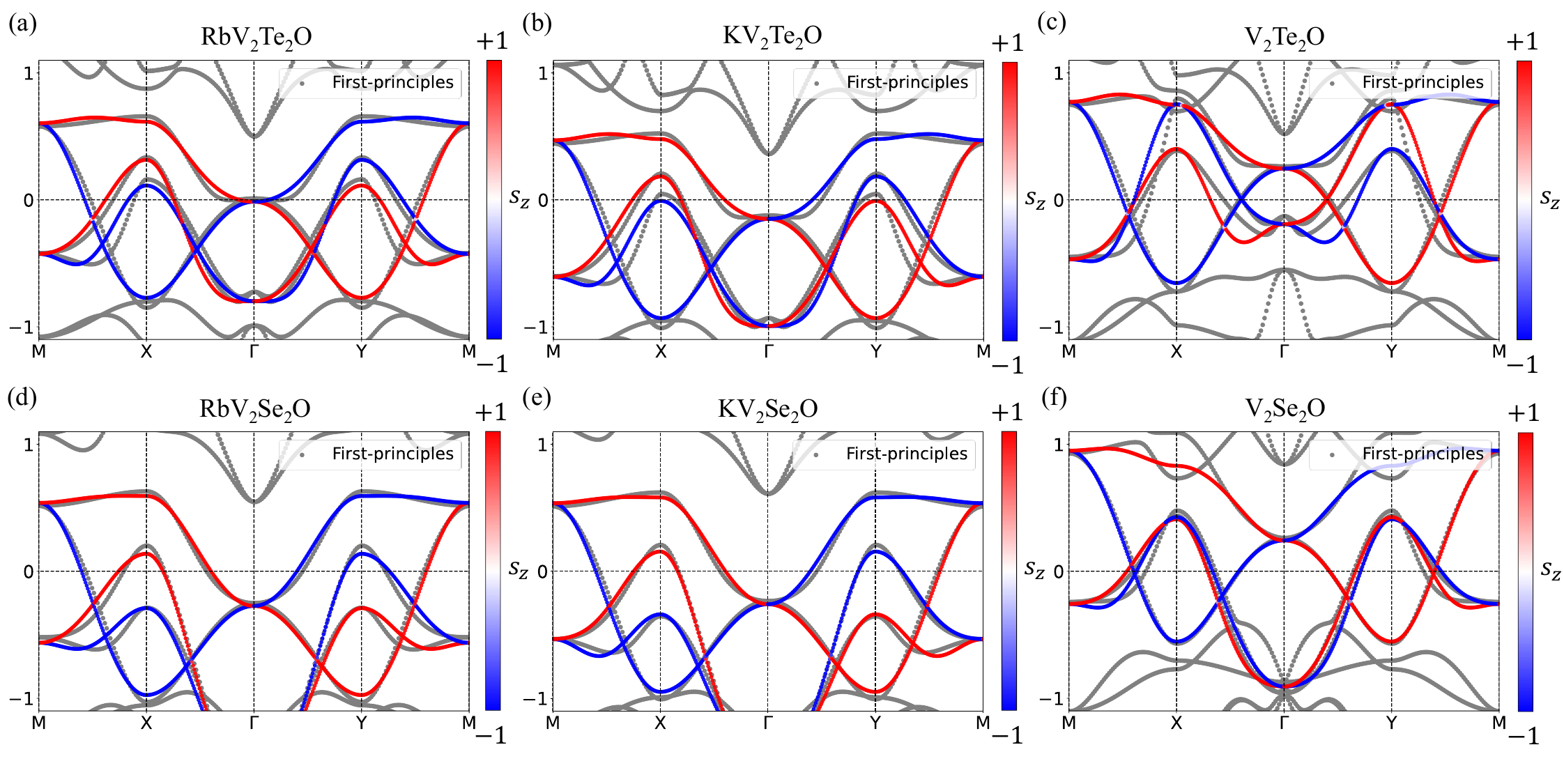}
	\caption{{\bf CSML band structures in the presence of SOC.} Band structures and the corresponding spin polarization (\(S_z\)) obtained from the TB model are represented by red and blue dots, respectively. Results from first-principles calculations are shown as grey dots for comparison.}
	\label{fig:soc}
\end{figure*}

After constructing the model, we fit all hopping parameters to match the first-principles calculations, which show excellent agreement with experimental data in Rb-intercalated V$_2$Te$_2$O and K-intercalated V$_2$Se$_2$O \cite{rvto_2025,kvso_2025}. Motivated by this correspondence between calculations and experiments, we employ first-principles calculations to obtain band structures for other members in the V$_2$Se$_2$O-family. The fitted hopping parameters are listed in Table \ref{parameter_nsoc} and a direct comparison between band structures from TB model and first-principles calculations is shown in Fig. \ref{fig:nsoc}.

\subsection{TB model with SOC}
To account for relativistic effects in V$_2$X$_2$O, we incorporate SOC into the TB model. The presence of SOC fundamentally couples spin and real space, enabling crystal symmetry to simultaneously constrain both spin and orbital degrees of freedom. This necessitates an analysis through the magnetic space group (MSG) framework to determine symmetry-allowed SOC terms. For V$_2$X$_2$O with the N\'eel vector aligned along the out-of-plane ($c$-axis) direction, the relevant MSG is  P4$^{\prime}$/mm$^{\prime}$m. The full TB Hamiltonian with SOC follows $H=H_0+H_{\bf{soc}}$, where $H_0$ denotes the Hamiltonian without SOC discussed previously, and $H_{\bf{soc}}$ is the symmetry-allowed SOC term, introducing finite interaction between spin-up and spin-down blocks, which is expressed as:

\begin{align}
H_{\bf{soc}} &= \begin{pmatrix}
0 & H_{\uparrow\downarrow} \\
H_{\uparrow\downarrow}^{\dagger} & 0
\end{pmatrix},\\
H_{\uparrow\downarrow} &= \begin{pmatrix}
0 & h_{0} \\
h_{1} & 0
\end{pmatrix},
\end{align}
with the explicit matrix elements $h_0$ and $h_1$ given in Appendix~\ref{appendix_hopping}.

\begin{figure}[hb]
	\centering
	\includegraphics[width=0.9\linewidth]{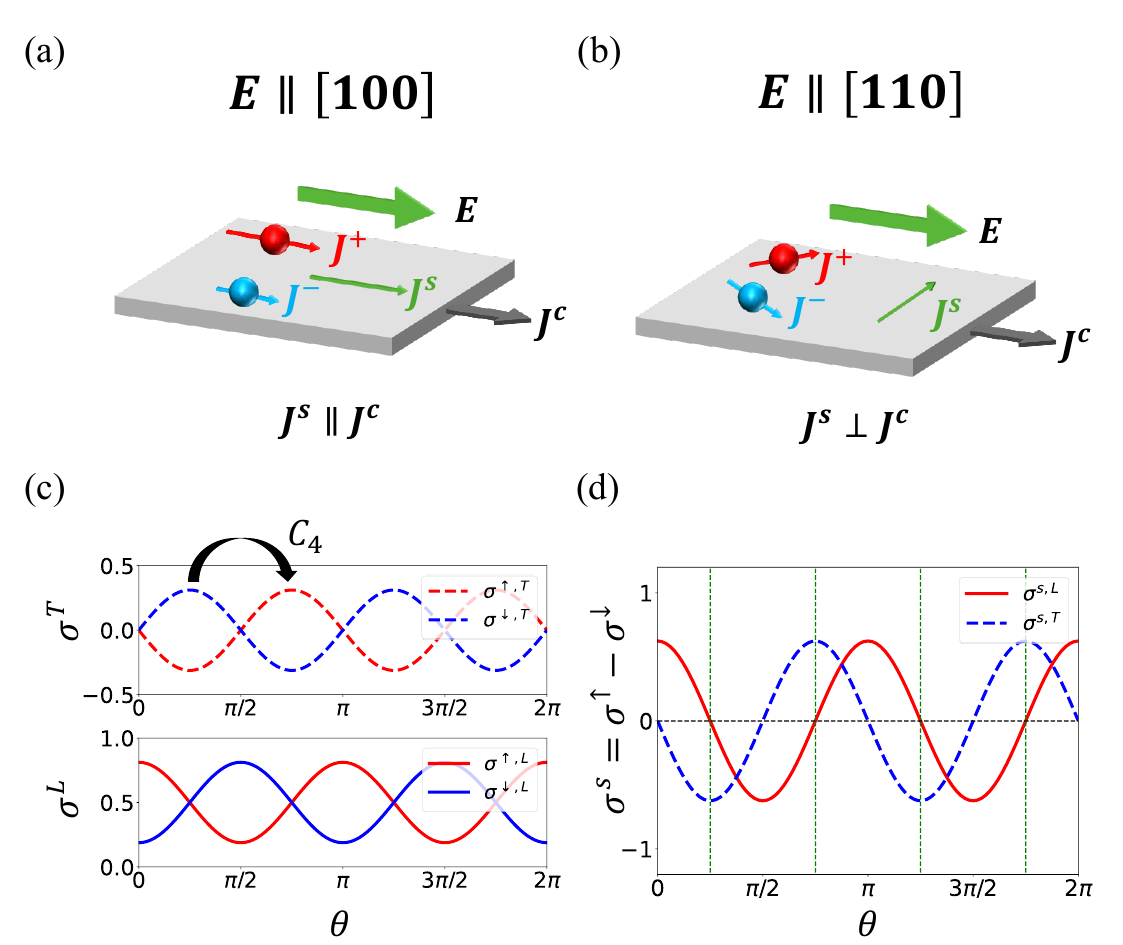}
	\caption{{\bf Noncollinear spin current in V$_2$X$_2$O.} (a) Spin polarized current ($\boldsymbol{J^s}\parallel \boldsymbol{J^c}$) when $\boldsymbol{E}\parallel[100]$. (b) Pure spin current ($\boldsymbol{J^s}\perp \boldsymbol{J^c}$) when $\boldsymbol{E}\parallel[110]$. (c) Angle-dependent longitudinal ($\sigma^{L}$) and transverse ($\sigma^{T}$) conductivity for each spin channel, normalized by $\sigma^{\uparrow,L}(\theta=0)+\sigma^{\downarrow,L}(\theta=0)$. (d) Corresponding angle-dependent longitudinal ($\sigma^{s,L}$) and transverse ($\sigma^{s,T}$) spin conductivities. Green dashed lines mark the field angles yielding a pure spin current ($\sigma^{s,L}=0$)}
	\label{fig:spin_current}
\end{figure}

In V$_2$X$_2$O, the spin‑splitting arises from a strong AFM exchange interaction that is much larger than the SOC energy scale. Consequently, the overall spin‑split bands exhibit only minor modifications when SOC is included, as shown in Fig.~\ref{fig:soc}. However, SOC introduces coupling between spin-up and spin-down blocks, opening finite gaps at originally spin-degenerated points. As a result, it generates sizable Berry curvature in the vicinity of these crossing points, allowing for the tunable Hall effect, for example, by strain as discussed in Sec.~\ref{strains}. 

We also fit the TB model including SOC term to the bands from first-principles calculations. The fitted parameters and corresponding spin-polarized band structures are given in Table \ref{parameter_soc} and Fig. \ref{fig:soc}, respectively. It is important to note that although SOC breaks spin conservation (where spin is no longer a good quantum number but a pseudo-vector), the magnetic crystal symmetries of the MSG still enforce the essential CSML pattern. For example, spin polarization \(S_z\) exhibits opposite signs in the $\mathcal{C}$-paired X and Y valleys (Fig.~\ref{fig:soc}), preserving the hallmark signature of spin–momentum locking even in the presence of SOC.

\begin{figure*}[tp]
	\centering
	\includegraphics[width=0.9\linewidth]{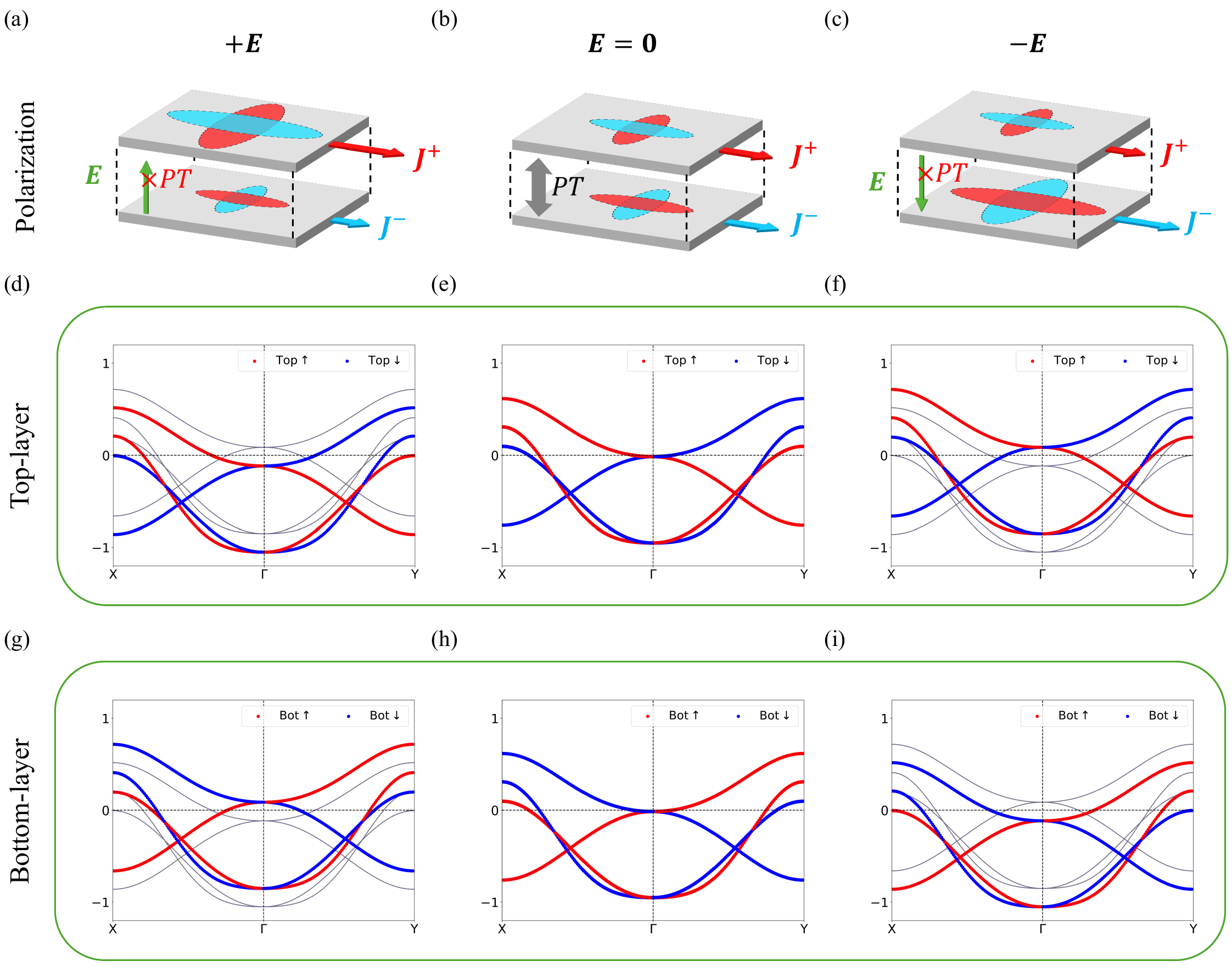}
	\caption{{\bf Electric-field control of layer polarization} (a)–(c) Schematic illustration of the induced layer polarization and the corresponding non‑relativistic spin current. (d)–(f) Band projection onto the top layer. (g)–(i) Band projection onto the bottom layer. The three columns correspond to an electric field oriented along $+z$, zero, and $-z$, respectively.}
	\label{fig:bilayer}
\end{figure*}

\section{Responses Under Electric Fields}\label{efield}
\subsection{Noncollinear spin currents}

Altermagnets uniquely enable noncollinear spin-conserved currents even with zero net magnetization, which is attributed to anisotropic conductivity connected by the crystal symmetry $\mathcal{C}$ \cite{maMultifunctional2021a,wuFermiLiquidInstabilities2007a,naka_spin_2019,gonzalez-hernandezEfficientElectricalSpin2021a}. Moreover, the orientation of this novel spin current depends on the direction of the applied electric field, manifesting in two distinct types, including both the spin-polarized currents similar to those in ferromagnets, arising from unequal velocities of spin-up and spin-down carriers (Fig. \ref{fig:spin_current}(a)); and the pure spin currents, which are analogous to the spin Hall effect but originate from anisotropic conductivity (Fig. \ref{fig:spin_current}(b)). This ability to generate these electric-field-tuned spin currents in the absence of SOC and net magnetization establishes altermagnets as an ideal platform for spin sources.

Here we simulate the nonrelativistic spin current in Rb-intercalated V$_2$Te$_2$O using our TB model. Due to the nonrelativistic nature of CSML, spin remains a good quantum number. This permits the separate definition and calculation for spin-polarized conductivities ($\sigma^{\uparrow}$ and $\sigma^{\downarrow}$) using the semiclassical Boltzmann equation \cite{solid_physics}. Figure \ref{fig:spin_current}(c) displays the calculated longitudinal ($\sigma^L$) and transverse ($\sigma^T$) components for each spin channel. Both components exhibit a pronounced angular dependence, revealing the strong anisotropy in the spin conductivity. Notably, $\sigma^{\uparrow,T(L)}$ and $\sigma^{\downarrow,T(L)}$ share identical forms versus electric field angle $\theta$ but exhibit a $\pi/2$ phase shift, which is a direct consequence of $\mathcal{C}_4$ symmetry-imposed spin-momentum locking in reciprocal space.

Correspondingly, the net spin conductivity $\sigma^s = \sigma^{\uparrow} - \sigma^{\downarrow}$ exhibits analogous angular dependence. Here, when $\boldsymbol{E} \parallel [100]$ or $[010]$ ($\theta = \frac{\pi}{2}N$), transverse components vanish ($\sigma^{\uparrow,T} = \sigma^{\downarrow,T} = 0$) yielding a longitudinal spin-polarized current (Fig. \ref{fig:spin_current}(a)). In contrast, when $\boldsymbol{E} \parallel [110]$ or $[1\bar{1}0]$ ($\theta = \frac{\pi}{2}N + \frac{\pi}{4}$), the longitudinal components become equal ($\sigma^{\uparrow,L} = \sigma^{\downarrow,L}$) while the the unequal transverse components produce a pure spin current (Fig. \ref{fig:spin_current}(b)). The complete angular dependence of $\sigma^{s,L}$ and $\sigma^{s,T}$ is summarized in Fig. \ref{fig:spin_current}(d).

\subsection{Layer-polarization by perpendicular electric field}

In low-dimensional systems, the layer degree of freedom serves as a crucial tunable index for field modulation and device applications\cite{layer_pesinSpintronics2012,layer_xuSpin2014}. When the layered index is further coupled with CSML, it gives rise to novel physical responses under external fields \cite{electric_2025}. Here we focus on the bilayer V$_2$Se$_2$O and V$_2$Te$_2$O as representative examples. We first determine the ground state of this bilayer system by evaluating different stacking orders and magnetic configurations. As detailed by the energy calculations in Appendix \ref{appendix_stacking}, this system favors AB stacking with intralayer AFM order. However, the weak interlayer coupling leads to only minimal energy differences (around meV) between FM and AFM interlayer aligments, leaving the precise interlayer magnetic order to be clarified by future experiments.

Crucially, these bilayer configurations can support both traditional spin-degenerate AFM state and spin-splitting CSML state (Table \ref{splitting}). While the CSML states directly produce noncollinear spin‑conserved currents under an in‑plane electric field (as discussed in the last subsection), here we investigate the layer‑polarization induced by an out-of-plane electric field in the spin‑degenerate AFM state, corresponding to AB stacking with intra‑layer AFM and inter‑layer FM order. For such a spin-degenerate system, the Hamiltonian is described by an expanded basis: \{$d_{xz}^{A_{bot}, \uparrow}, d_{xy}^{A_{bot},\uparrow}, d_{yz}^{B_{bot}, \downarrow}, d_{xy}^{B_{bot}, \downarrow},$ $ d_{xz}^{A_{top}, \downarrow}, d_{xy}^{A_{top},\downarrow}, d_{yz}^{B_{top}, \uparrow}, d_{xy}^{B_{top}, \uparrow}$\}, and the corresponding Hamiltonian can be written as:
\begin{equation}
    H_{B} = \begin{pmatrix}
        H_0 + \alpha E & H_{c} \\
        H_{c}^{\dagger}  & H_0 - \alpha E
    \end{pmatrix},
\end{equation}
where $H_0$ represents the single-layer Hamiltonian discussed above, $H_{c}$ describes interlayer coupling (see Appendix. \ref{appendix_bilayer} for details), and $\alpha E$ quantifies the electric field response. 

Band structures calculated from TB models under positive, zero and negative perpendicular electric field are shown in Fig.~\ref{fig:bilayer}. In the absence of electric field, $\mathcal{PT}$ symmetry ensures the identical dispersions of the top and bottom layers but with reversed spin polarization (Fig. \ref{fig:bilayer}(e) and (h)), resulting in full spin-degeneracy throughout the Brillouin zone. However, when a perpendicular electric field is applied, it breaks $\mathcal{PT}$ symmetry, introducing finite layer polarization while preserving the CSML pattern (Fig.~\ref{fig:bilayer}(d) and (g)). Crucially, reversing the direction of perpendicular field inverts both layer polarization and related spin polarization (Fig.~\ref{fig:bilayer}(f) and (i)), thus enabling simultaneous electrical control of layer and spin configurations.

A similar analysis applies to the nonrelativistic spin currents (Fig. \ref{fig:bilayer}(a)-(c)). At zero electric field, the net spin current vanishes due to exact cancellation between the spin current of top and bottom layers carrying opposite spin. Once a perpendicular field breaks $\mathcal{PT}$ symmetry, it leads to finite layer-polarization as discussed above, generating finite net spin current (Fig.~\ref{fig:bilayer}(a)). Crucially, the polarization of spin current is reversed when the perpendicular field is flipped, following the switched layer-polarization. Therefore, this mechanism enables electrical control of spin responses, establishing a critical platform for next-generation spintronic memory and logic devices via purely electrical switching.

\section{Strain modulated effects}\label{strains}

\subsection{Strain-dependent modification of the TB model}

\begin{figure}[b!]
	\centering
	\includegraphics[width=0.9\linewidth]{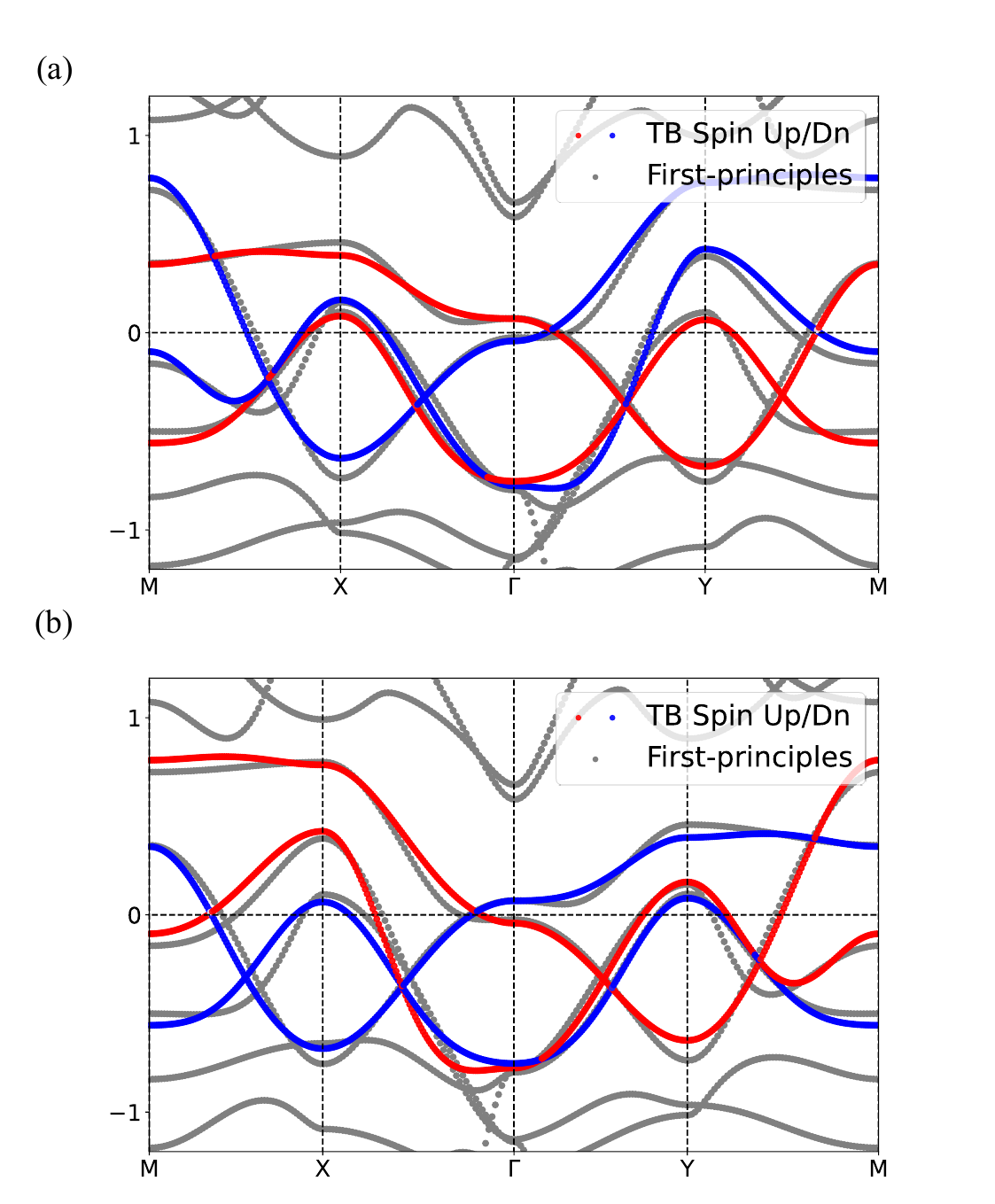}
	\caption{{\bf Strained band structures.} Band structures under (a) $\epsilon_{xx}=5\%$ and (b) $\epsilon_{yy}=5\%$ uniaxial strain. Bands calculated by TB model are denoted by red (blue) dots for spin-up (spin-down), first-principles calculation results are denoted by grey dots.}
	\label{fig:strain}
\end{figure}

Strain engineering provides a powerful approach for tuning material properties, enabling effective modulation of electronic structures in low‑dimensional materials \cite{strain_zhaoStrain2020}. In altermagnets, such modifications can lead to novel phenomena like unconventional piezomagnetism \cite{maMultifunctional2021a,KB_strain2025}, strain-induced spin-splitting torque \cite{KB_SST_2025}, and anomalous Hall effect \cite{zhou_crsb_2025}. To systematically investigate strain effects in V$_2$X$_2$O, we develop a comprehensive TB model incorporating strain-dependent modifications to hopping parameters.

Under applied strain tensor $\epsilon$, lattice vectors deform as:

\begin{equation}
	\vec{a}_i \rightarrow (I + \epsilon) \cdot \vec{a}_i,
\end{equation}
modifying hopping parameters through:
\begin{equation}
	t_{ij}(\epsilon) = t_{ij}^0 + \sum_{\alpha,\beta} \frac{\partial t_{ij}}{\partial \epsilon_{\alpha\beta}}\epsilon_{\alpha\beta} = t_{ij}^0 + \sum_{\alpha,\beta} \Lambda_{\alpha\beta}\epsilon_{\alpha\beta},
\end{equation}
where $t_{ij}^0$ denotes unstrained hoppings and $\Lambda_{\alpha\beta}$ represents strain-dependent coefficients.

Here, strain induces two distinct types of modification to the TB model: (1) symmetry‑preserving adjustments that uniformly alter the band dispersion without lowering the original crystal symmetry; and (2) symmetry‑breaking terms that lift degeneracies between equivalent hopping. Accordingly, the strained Hamiltonian can be decomposed as:
\begin{equation}
	H (\epsilon)= H_0 + H^{\Lambda^0} \cdot \epsilon + H^{\Lambda^{\prime}} \cdot \epsilon,
\end{equation}
where $H^{\Lambda^0}$ denotes the symmetry‑preserving corrections and $H^{\Lambda^{\prime}}$ governs the symmetry‑breaking contributions responsible for unconventional piezomagnetism and piezo‑Hall effect. The explicit forms of these terms are provided in Appendix~\ref{appendix_strain}. Then we fit the strain-related coefficients by the first-principles calculations, which values are shown in the Table \ref{parameter_strain}. 

Our parameterized model quantitatively reproduces the first-principles calculations under strain (Fig. \ref{fig:strain}), provides a powerful framework for investigating how mechanical deformations modify the electronic properties of V$_2$X$_2$O systems. This enables quantitative exploration of strain-modulated phenomena including band structure modulation, tunable spin splitting, unconventional piezomagnetism and piezo-Hall effects \cite{maMultifunctional2021a,hu_2025}.

\subsection{Strain modulated Berry curvature}
\begin{figure}[htbp]
    \centering
    \includegraphics[width=0.9\linewidth]{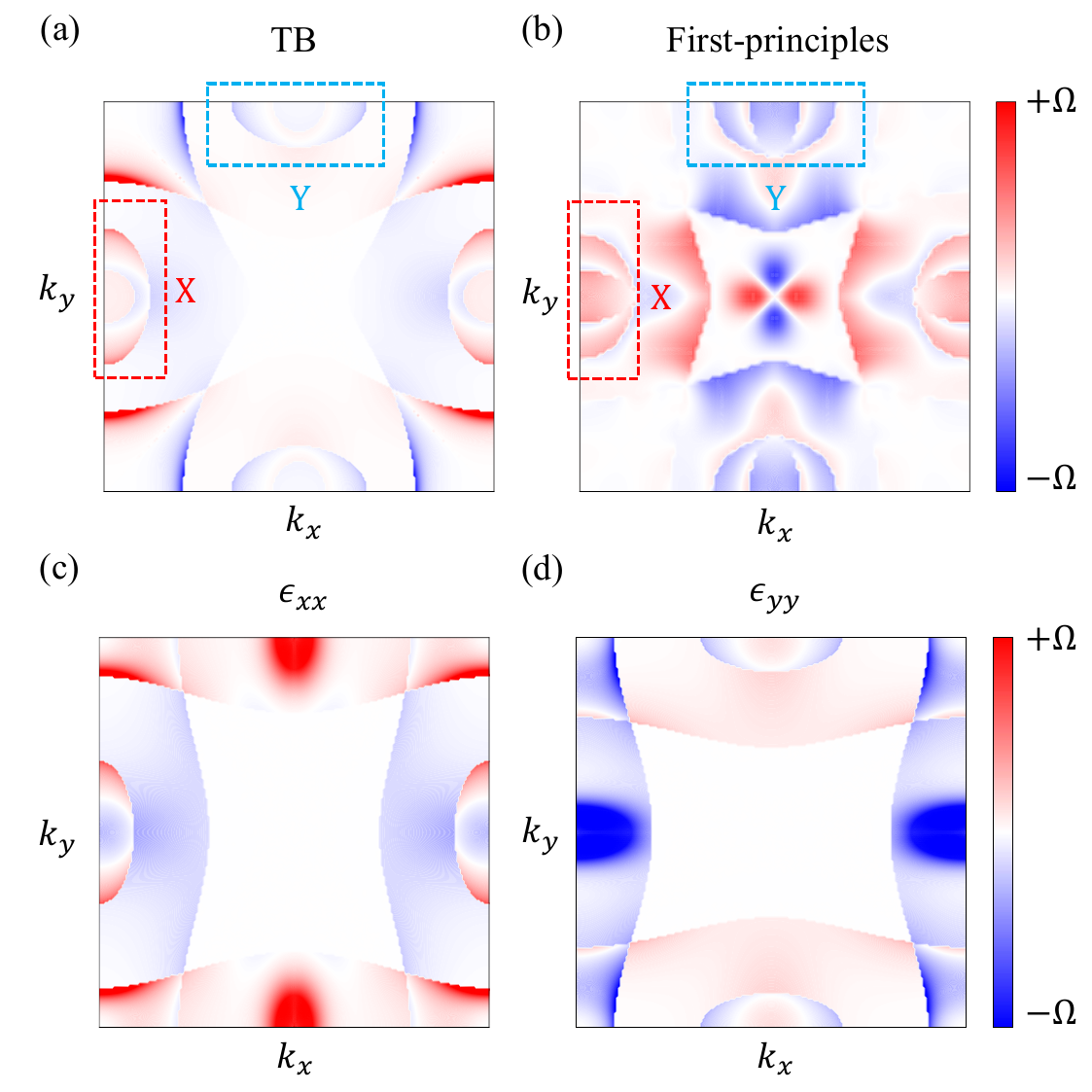}
    \caption{{\bf Berry curvature from TB model vs. first-principles calculations.} (a) Berry curvature component $\Omega_z$ computed with the TB model (no strain), showing opposite signs in the $\mathcal{C}$-paired X and Y valleys. (b) Corresponding first-principles calculation result; the TB model agrees well except in the central region, where deep bands below the Fermi level contribute. (c, d) When a symmetry‑breaking strain tensor component $\epsilon_{xx}$ or $\epsilon_{yy}$ is applied, $\Omega_z$ no longer cancels exactly between $\mathcal{C}$-paired momenta, giving rise to a finite anomalous Hall conductivity.}
    \label{fig:strained_BC}
\end{figure}

Here we calculate the distribution of Berry curvature (Fig.~\ref{fig:strained_BC}(a)). The results match well with the first-principles calculations (Fig.~\ref{fig:strained_BC}(b)), except the central region originated from bands far away from the Fermi level (near –1 eV in Fig.~\ref{fig:soc}), which can be captured by including more orbitals. Importantly, in the unstrained crystal, the P4$^\prime$/mm$^\prime$m magnetic space group enforces a strict cancellation of Berry curvature between symmetry‑related momenta:
\begin{equation}
	\Omega_z(\mathbf{k}) = -\Omega_z(\mathcal{C}\mathbf{k})
\end{equation}
where $\mathcal{C}$ denotes the symmetry operations that connect the X and Y valleys (Fig. \ref{fig:strained_BC}(a)). This symmetry protection prohibits both net magnetization and anomalous Hall conductivity. 

However, application of $\epsilon_{xx}$ or $\epsilon_{yy}$ strain fundamentally disrupts this balance: the strain breaks MSG symmetries (like $\mathcal{C}_{4z}$ rotation and mirror $\mathcal{M}_{\bar{x}y}$) connecting opposite sublattices, lowering the magnetic space group to Pm$^\prime$m$^\prime$m and preserving mmm point group symmetry locally at each sublattice. Consequently, Berry curvature no longer cancels between formerly equivalent $\mathcal{C}$-paired momentum points (Figs. \ref{fig:strained_BC} (c) and (d)), yielding finite anomalous Hall conductivity. 

\section{Summary}\label{summary}

In this work, we develop a realistic tight-binding model for layered V$_2$Se$_2$O-family $d$‑wave altermagnets, which can also be used for other similar compounds \cite{NSR2025}. Parameterized from first‑principles calculations, this four‑band model accurately captures the crystal-symmetry paired electronic structures, both with and without spin–orbit coupling. Beyond reproducing the band dispersion, it provides a versatile platform to simulate novel phenomena such as noncollinear spin‑conserved currents, layer polarization controlled by an out‑of‑plane electric field, and strain‑tunable piezo‑Hall effects. Furthermore, the model can be extended to simulate heterostructures that integrate V$_2$Se$_2$O-family with other quantum materials—such as superconductors, topological insulators, or two‑dimensional magnets—offering a theoretical framework to study proximity‑induced phenomena, interfacial effects, and emergent hybrid functionalities. Furthermore, it reveals potential external‑field modulation of low‑dimensional altermagnetism via strain, electric field, and magnetic field. Therefore, this realistic TB framework provides a predictive tool for exploring coupling effects among multiple degrees of freedom such as spin, valley, and layer, establishing a solid theoretical foundation for understanding layered altermagnetism and guiding the design of future spintronic devices based on V$_2$Se$_2$O-family.

\section*{Acknowledgments}
This work is supported by National Key R$\&$D Program of China (2021YFA1401500) and the Hong Kong Research Grants Council (CRS-HKUST603/25, C6046-24G, 16306722, 16304523 and 16311125).

\onecolumngrid
\clearpage
\section*{Appendix}\label{appendix}

\subsection{Band projections for V$_2$X$_2$O family}\label{appendix_projection}

\begin{figure*}[ht]
	\centering
	\includegraphics[width=0.9\linewidth]{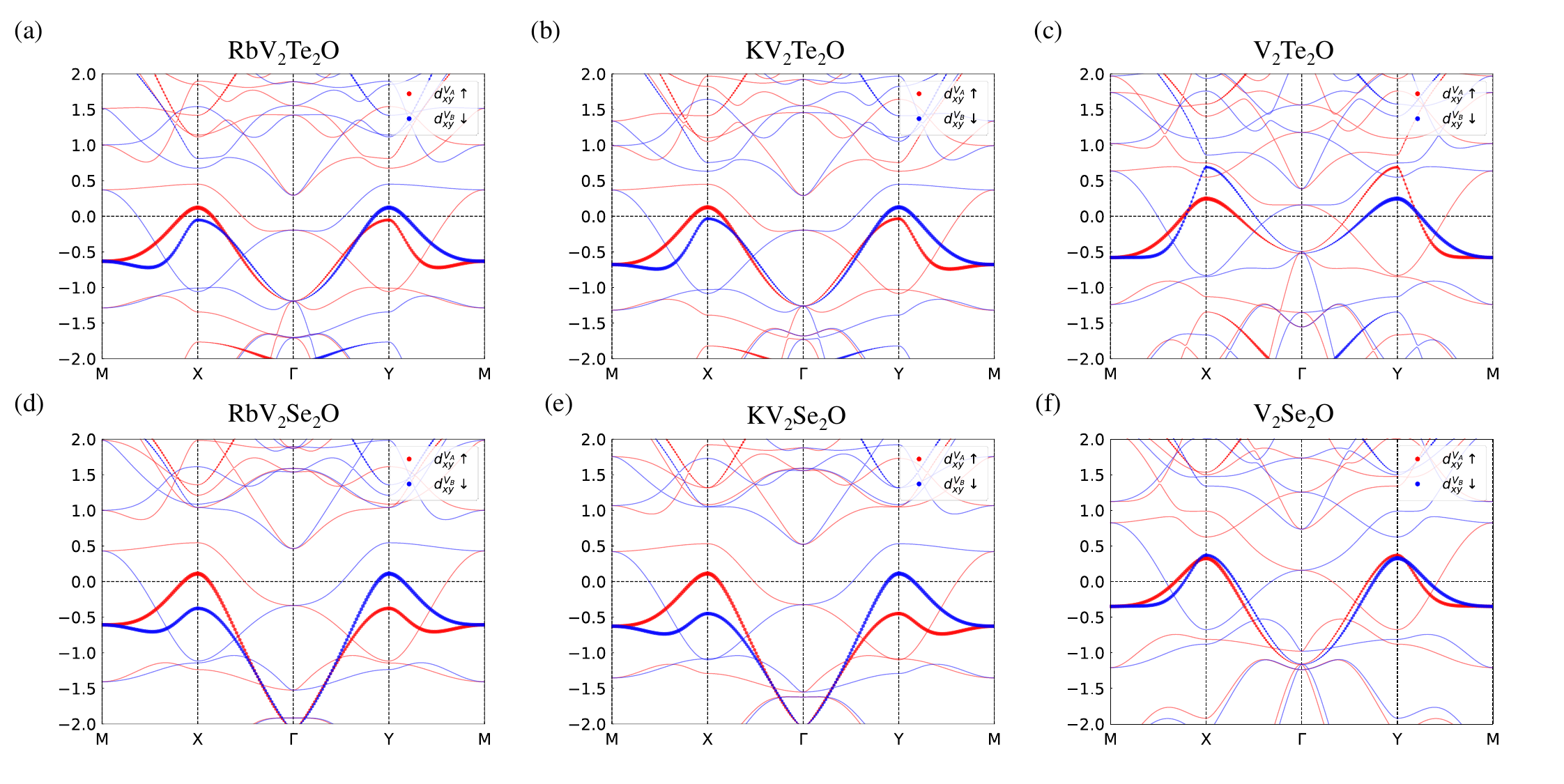}
	\caption{{\bf Orbital projections of $d_{xy}$ for V$_2$X$_2$O family.} The size of the colored circles is proportional to the weight of projection onto atomic orbitals $d_{xy}^{\uparrow}$ from V$_A$ and  $d_{xy}^{\downarrow}$ from V$_B$. Here red and blue circles refer to spin-up and spin-down, respectively.}
	\label{fig:proj_xy}
\end{figure*}

\begin{figure*}[ht]
	\centering
	\includegraphics[width=0.9\linewidth]{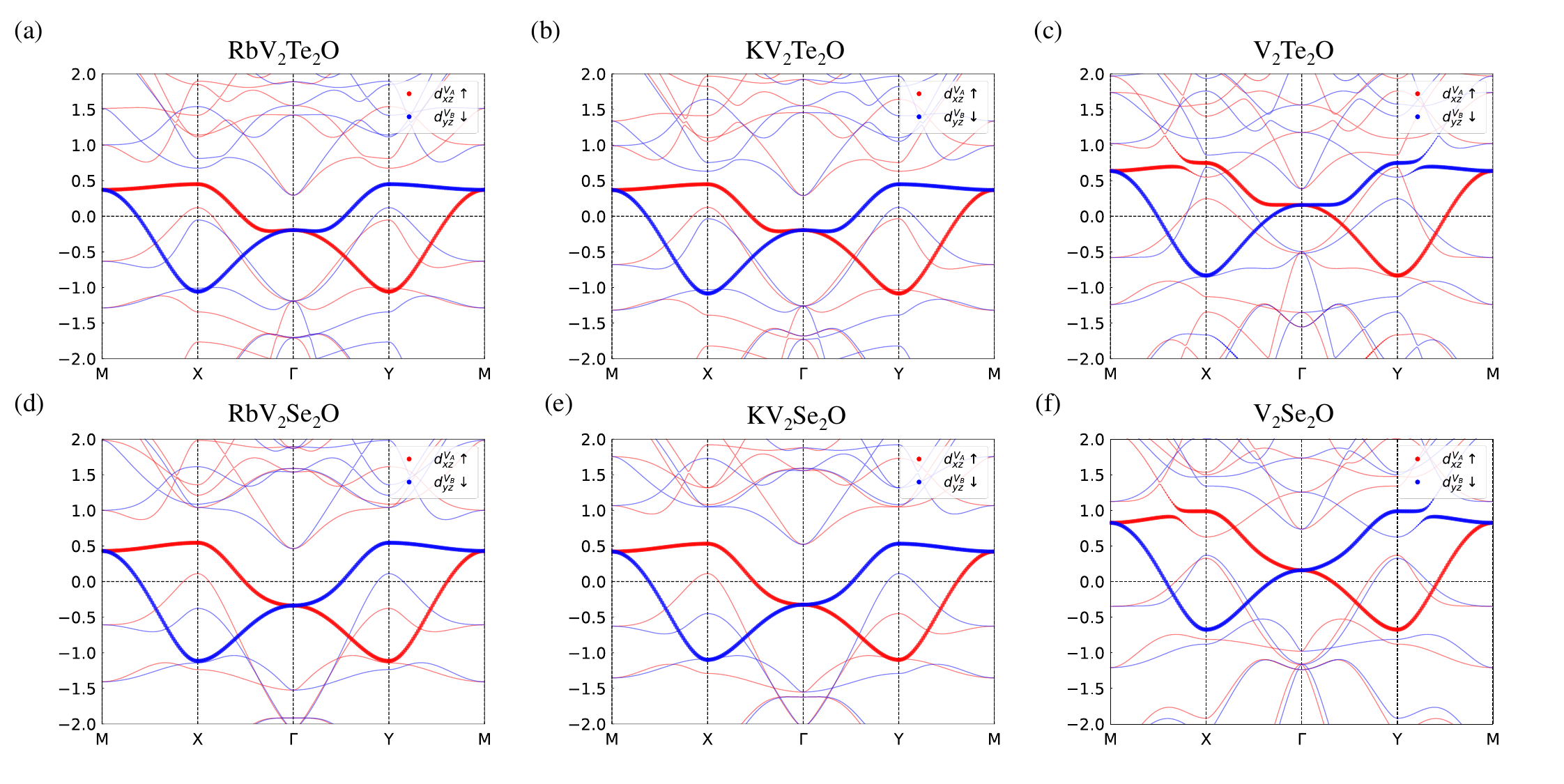}
	\caption{{\bf Orbital projections of $d_{xz}/d_{yz}$ for V$_2$X$_2$O family.} The size of the colored circles is proportional to the weight of projection onto atomic orbitals $d_{xz}^{\uparrow}$ from V$_A$ and $d_{yz}^{\downarrow}$ from V$_B$. Here red and blue circles refer to spin-up and spin-down, respectively.}
	\label{fig:proj_xz}
\end{figure*}

\clearpage

\subsection{Band structures of V$_2$Se$_2$O and V$_2$Te$_2$O for different values of U}

\begin{figure*}[ht]
	\centering
	\includegraphics[width=0.9\textwidth]{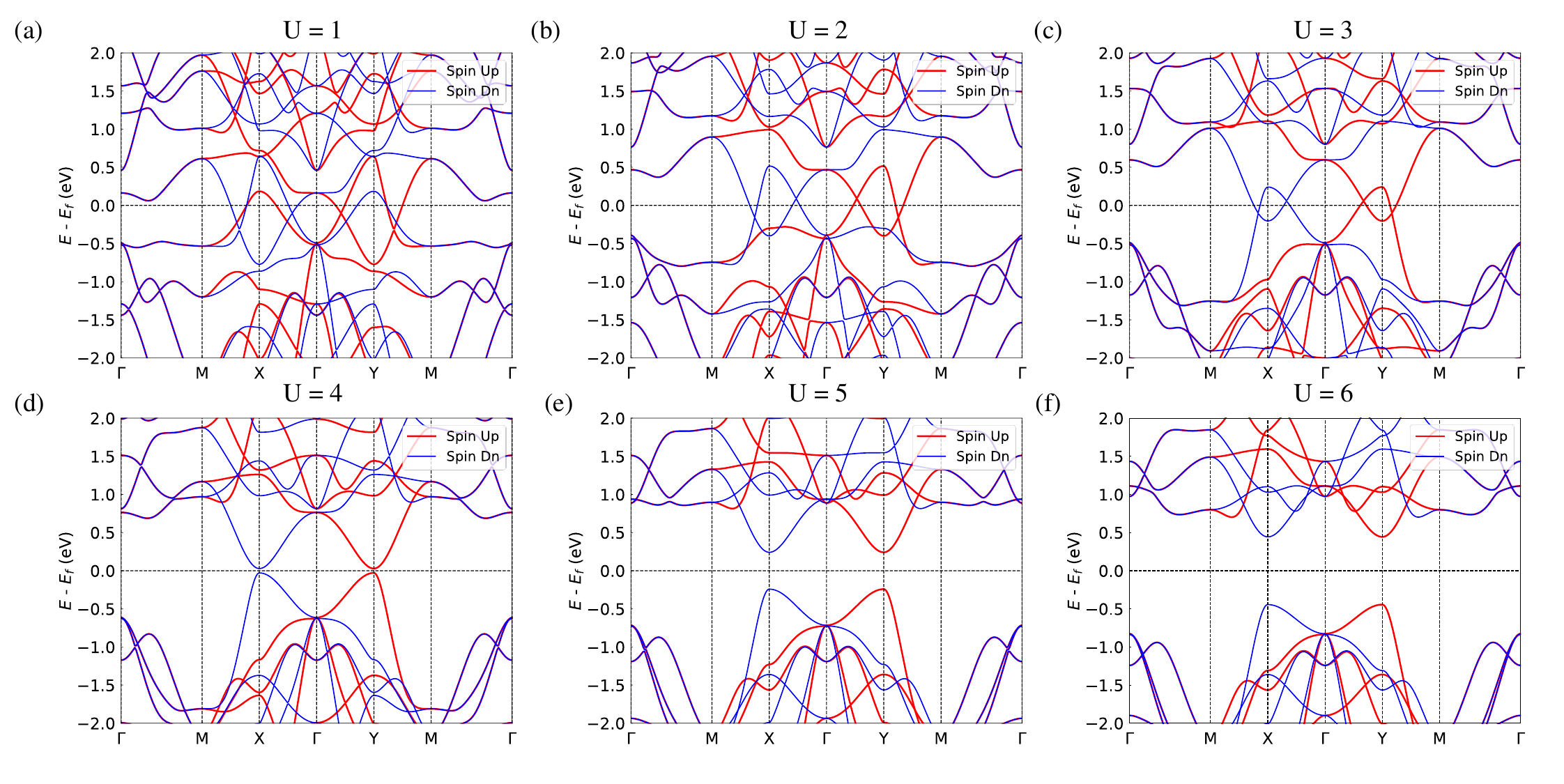}
	\caption{{\bf First-principles calculations for monolayer V$_2$Te$_2$O with different U}}
	\label{fig:VTO_U}
\end{figure*}

\begin{figure*}[ht]
	\centering
	\includegraphics[width=0.9\textwidth]{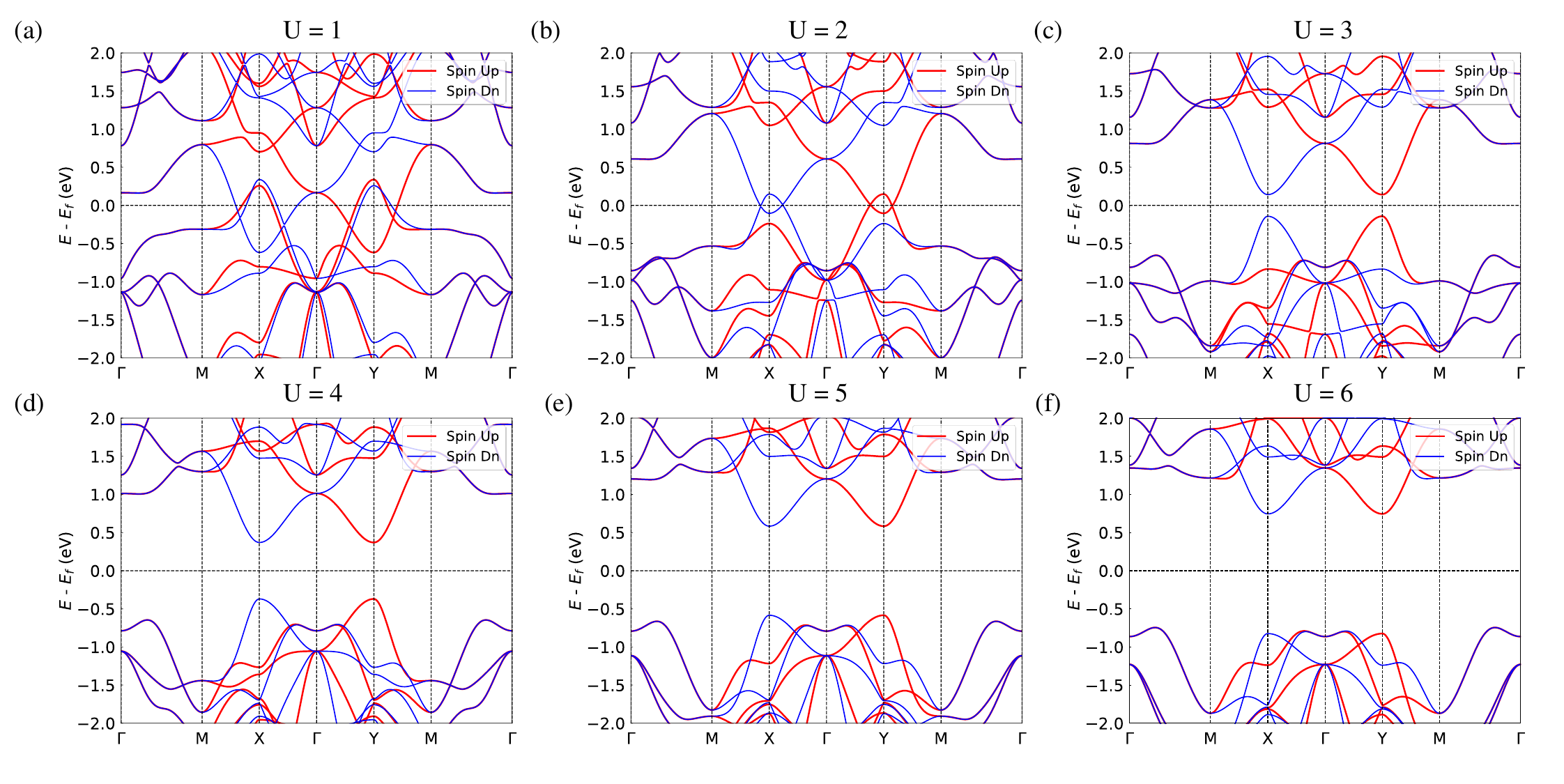}
	\caption{{\bf First-principles calculations for monolayer V$_2$Se$_2$O with different U}}
	\label{fig:VSO_U}
\end{figure*}

\clearpage

\subsection{Hopping terms and symmetry constraints}\label{appendix_hopping}
\begin{figure}[ht]
	\centering
	\includegraphics[width=0.9\textwidth]{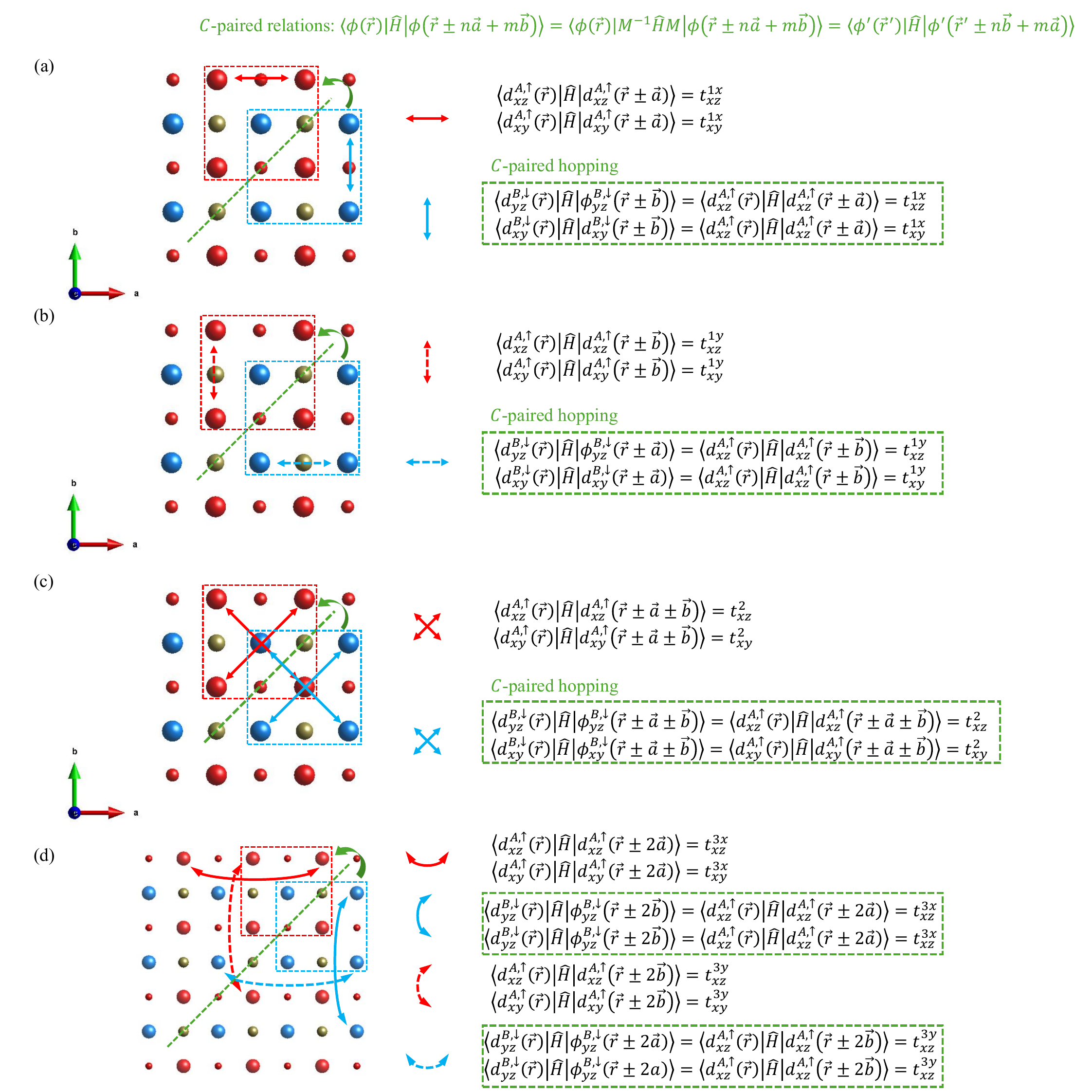}
	\caption{{\bf Symmetry-paired hopping terms in real space for V$_2$X$_2$O TB model.} }
	\label{fig:hopping}
\end{figure}

The crystal symmetry $\mathcal{C}$, which connects antiferromagnetic sublattices in real space and gives rise to $\mathcal{C}$-paired spin–momentum locking in reciprocal space, imposes equivalent relations among symmetry-paired hopping terms. Formally, a hopping integral is defined as $t_{ij} = \langle \phi(\mathbf{r}_i) | \hat{H} | \phi(\mathbf{r}_j) \rangle$, where $\phi(\mathbf{r}_i)$ and $\phi(\mathbf{r}_j)$ denote orbital basis states at lattice sites $i$ and $j$, respectively. Symmetry invariant relation requires that: $\langle \phi(\mathbf{r}_i) | \hat{H} | \phi(\mathbf{r}_j) \rangle = \langle \phi(\mathbf{r}_i) | \mathcal{C}^{-1} \hat{H} \mathcal{C} | \phi(\mathbf{r}_j) \rangle$, linking equivalent hopping in real space. As an explicit example, under the mirror operation $\mathcal{M}_{\bar{x}y}$ that exchanges the A and B sublattices (Fig.~\ref{fig:hopping}), the orbital transformation $d_{xz}^{A,\uparrow} \rightarrow d_{yz}^{B,\downarrow}$ leads to the equivalence:
$\langle d_{xz}^{A,\uparrow}(\mathbf{r}) | \hat{H} | d_{xz}^{A,\uparrow}(\mathbf{r} + a\hat{x}) \rangle = \langle d_{yz}^{B,\downarrow}(\mathbf{r}) | \hat{H} | d_{yz}^{B,\downarrow}(\mathbf{r} + b\hat{y}) \rangle$. Thus, an $x$-direction hop on sublattice A becomes symmetry-equivalent to a $y$-direction hop on sublattice B. All symmetry-constrained hopping equivalences are summarized schematically in Fig.~\ref{fig:hopping}.

\clearpage

Therefore, the symmetry-restricted TB model without SOC is expressed as:

\begin{equation}
H_{0}= \begin{pmatrix}
H_{xz, \uparrow} & 0 & 0 & 0 \\
0 & H_{xy, \uparrow} & 0 & 0 \\
0 & 0 & H_{y z, \downarrow} & 0 \\
0 & 0 & 0 & H_{xy, \downarrow}
\end{pmatrix},
\end{equation}
with the matrix elements given by:
\begin{align*}
H_{xz, \uparrow} & = e_{xz}+ 2t_{xz}^{1x} \cos k_x+2t_{xz}^{1y} \cos k_y+ 4 t_{xz}^{2} \cos k_x \cos k_y  + 2t_{xz}^{3x} \cos 2 k_x + 2t_{xz}^{3y} \cos 2 k_y, \\
H_{xy, \uparrow} & =  e_{xy}+ 2t_{xy}^{1x} \cos k_x+2t_{xy}^{1y} \cos k_y+4 t_{xy}^{2} \cos k_x \cos k_y  + 2t_{xy}^{3x} \cos 2 k_x + 2t_{xy}^{3y} \cos 2 k_y,  \\
H_{yz, \downarrow}&=e_{xz}+ 2 t_{xz}^{1x} \cos k_y+2 t_{xz}^{1y} \cos k_x+4 t_{xz}^{2} \cos k_x \cos k_y + 2t_{xz}^{3x} \cos 2 k_y + 2t_{xz}^{3y} \cos 2 k_x, \\
H_{x y, \downarrow}&=e_{xy}+ 2 t_{xy}^{1x} \cos k_y+2 t_{xy}^{1y} \cos k_x+4 t_{xy}^{2} \cos k_x \cos k_y + 2t_{xy}^{3x} \cos 2 k_y + 2t_{xy}^{3y} \cos 2 k_x.
\end{align*}

The SOC term is expressed as:

\begin{equation}
H_{\text{soc}}= \begin{pmatrix}
0 & 0 & 0 & h_{03} \\
0 & 0 & h_{12} & 0 \\
0 & h_{21} & 0 & 0 \\
h_{30} & 0 & 0 & 0
\end{pmatrix},
\end{equation}
with the matrix elements given by:

\begin{align*}
h_{03} =&4 i \lambda_1 \cos \frac{k_x}{2}\cos \frac{k_y}{2} -4 \lambda_2 \sin \frac{k_x}{2} \sin \frac{k_y}{2}, \\
h_{12} =&4 \lambda_1 \cos \frac{k_x}{2}\cos \frac{k_y}{2}-4 i \lambda_2 \sin \frac{k_x}{2} \sin \frac{k_y}{2}, \\
h_{21} =&4 \lambda_1 \cos \frac{k_x}{2}\cos \frac{k_y}{2}+4 i \lambda_2 \sin \frac{k_x}{2} \sin \frac{k_y}{2},\\
h_{30} =&-4 i \lambda_1 \cos \frac{k_x}{2}\cos \frac{k_y}{2}-4 \lambda_2 \sin \frac{k_x}{2} \sin \frac{k_y}{2}.
\end{align*}

\clearpage

\begin{table}[htbp!]
\setlength{\tabcolsep}{0.02\textwidth}
\begin{tabular}{lccccccc}
\hline
\hline
& $e_{xz}$ & $t_{xz}^{1x}$ & $t_{xz}^{1y}$ & $t_{xz}^{2}$ & $t_{xz}^{3x}$ & $t_{xz}^{3y}$ & $\lambda_1$ \\
& $e_{xy}$ & $t_{xy}^{1x}$ & $t_{xy}^{1y}$ &$ t_{xy}^{2}$ & $t_{xy}^{3x}$ & $t_{xy}^{3y}$ & $\lambda_2$ \\
\hline 
$\mathrm{RbV}_2\mathrm{Te}_2\mathrm{O}$& -0.04629229	& -0.24988718 &	0.094201	&0.04609694  & 0.01695743 &	-0.00710368 & -\\ 
&-0.54598791	&-0.08869994  &	-0.03574438	&-0.1130422	 & 0.06412028 &	0.01420836  & -\\
\hline 
$\mathrm{KV}_2\mathrm{Se}_2\mathrm{O}$&-0.02378745	&-0.29092429  &	0.09126757	&0.04004989	 & 0.00558104 &	-0.00366236 & -\\ 
&-0.71549718	&-0.2089599	  &-0.0856611	&-0.13369163 & 0.02983547 &	0.00226622  & -\\
\hline 
$\mathrm{RbV}_2\mathrm{Se}_2\mathrm{O}$&-0.03046053	&-0.2971818	  &0.09292557	&0.04011348	 & 0.00453865 &	-0.00244148 & -\\ 
&-0.71015906	&-0.20442841  &	-0.09841807	&-0.13850689 & 0.03057077 &	0.00361299  & -\\ 
\hline 
$\mathrm{KV}_2\mathrm{Te}_2\mathrm{O}$&-0.04658959	&-0.25303396  &	0.09691944	&0.04785715	 & 0.0171774  &	-0.00838878 & -\\ 
&-0.56526676	&-0.08930631  &	-0.03883836	&-0.12084957 & 0.06404315 &0.01571452   & -\\ 
\hline 
$\mathrm{V}_2\mathrm{Se}_2\mathrm{O}$&0.42248278&	-0.26138292&	0.0842136&	0.05722691&	-0.01128209	& -0.01455688& -\\ 
&-0.29163339&	-0.08229672&	-0.09272593&	-0.13273838	&0.07659586&	0.0292694   & -\\ 
\hline 
$\mathrm{V}_2\mathrm{Te}_2\mathrm{O}$&0.29332667&	-0.23516933&	0.10373498	&0.06058761&	0.010611&	-0.01618926 & -\\ 
&-0.10463117&	0.06942459&	-0.03189352&	-0.12280566&	0.05456965&	0.04649913     & -\\ 
\hline
\hline
\end{tabular}
\caption{{\bf Fitted TB model parameters for all six V$_2$X$_2$O in the absence of SOC.} All parameters are in units of eV.}
\label{parameter_nsoc}
\end{table}

\begin{table}[ht!]
\setlength{\tabcolsep}{0.015\textwidth}
\begin{tabular}{lccccccc}
\hline
\hline
& $e_{xz}$ & $t_{xz}^{1x}$ & $t_{xz}^{1y}$ & $t_{xz}^{2}$ & $t_{xz}^{3x}$ & $t_{xz}^{3y}$ & $\lambda_1$ \\
& $e_{xy}$ & $t_{xy}^{1x}$ & $t_{xy}^{1y}$ &$ t_{xy}^{2}$ & $t_{xy}^{3x}$ & $t_{xy}^{3y}$ & $\lambda_2$ \\
\hline 
$\mathrm{RbV}_2\mathrm{Te}_2\mathrm{O}$
&-0.05223832	&-0.25078344	&0.0961765	&0.04660479&	0.01985274&	-0.00953854&	0.00038615 \\
&-0.52960812	&-0.07193622	&-0.02131625	&-0.10328106&	0.0721981&	0.02303562&	0.00116775\\
\hline 
$\mathrm{KV}_2\mathrm{Se}_2\mathrm{O}$
&-0.02692227	&-0.29104709	&0.09206813	&0.04028453&	0.0067143&	-0.00456873&	0.00288065 \\
&-0.72424086	&-0.18424356	&-0.06059493	&-0.11643046&	0.055216&	0.02809002&	0.00003625\\
\hline 
$\mathrm{RbV}_2\mathrm{Se}_2\mathrm{O}$
&-0.03430049	&-0.2975763	    &0.09451738	&0.04049088&	0.00686343&	-0.00419609&	0.0030892 \\
&-0.70774909	&-0.19578962	&-0.08882642&	-0.13206917&	0.03876358&	0.0130189&	0.0001281\\
\hline 
$\mathrm{KV}_2\mathrm{Te}_2\mathrm{O}$
&-0.05168974	&-0.25373628	&0.0988842	&0.04824782&0.02020899&	-0.01108952&	0.00012773 \\
&-0.54823748	&-0.07236437	&-0.02433769&	-0.11104794&	0.07223957&	0.0236946&	0.00026778\\
\hline 
$\mathrm{V}_2\mathrm{Se}_2\mathrm{O}$
&0.41946488	&-0.26146334	&0.08463559	&0.05725641	&-0.0104565	&-0.01551344	&0.002328842\\
&-0.26985667&	-0.07928053&	-0.08329064&	-0.12516649&	0.06438239&	0.02954217&	0.001277906\\
\hline 
$\mathrm{V}_2\mathrm{Te}_2\mathrm{O}$
&0.279678	&-0.2412327	&0.11000072	&0.05778583	&0.01640664	&-0.01730426	&0.002132498 \\
&-0.13757969	&0.07875853	&-0.00944596	&-0.1133831	&0.0955085	&0.03527756	&0.001108951\\
\hline
\hline
\end{tabular}
\caption{{\bf Fitted TB model parameters for all six V$_2$X$_2$O in the presence of SOC.} All parameters are in units of eV.}
\label{parameter_soc}
\end{table}

\clearpage
\twocolumngrid
\subsection{Stacking orders of bilayer systems}\label{appendix_stacking}

\begin{figure}[ht]
	\centering
	\includegraphics[width=0.9\linewidth]{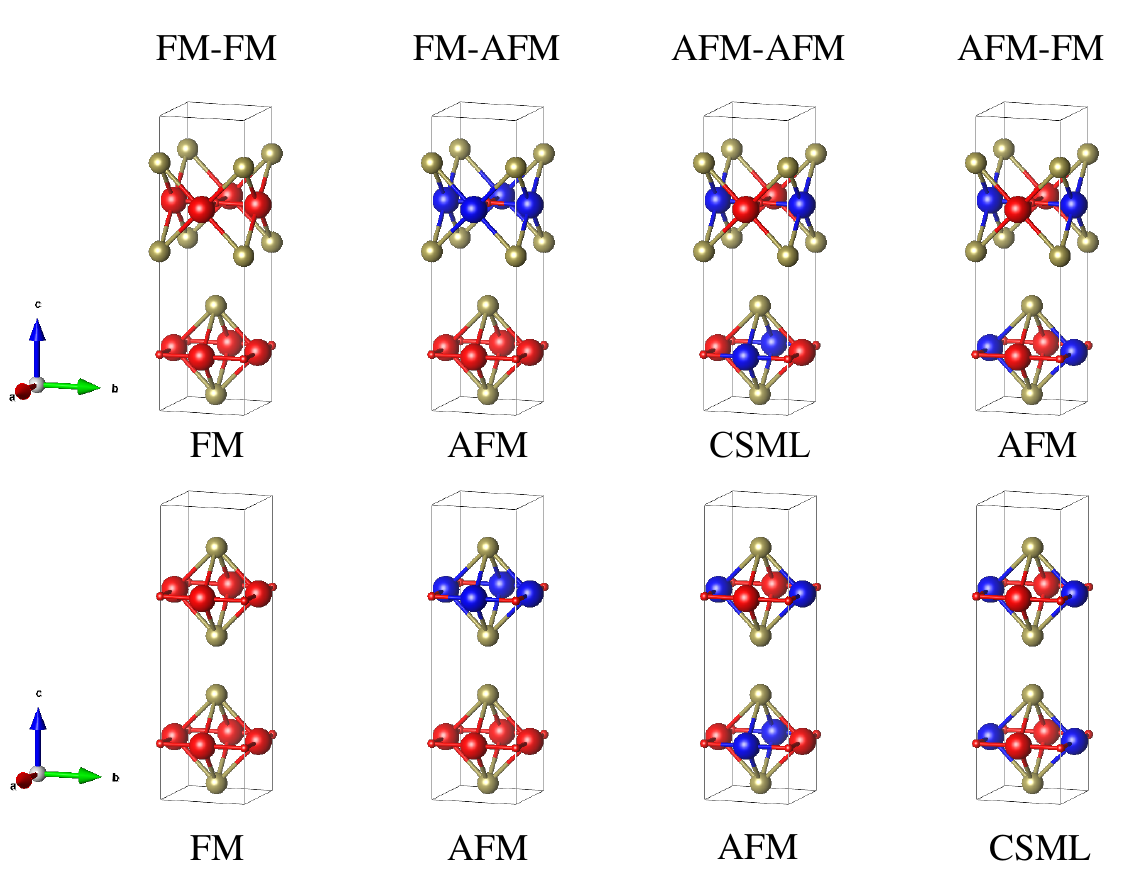}
	\caption{{\bf Different stacking orders of V$_2$X$_2$O family.} Here, FM-AFM refer to interlayer ferromagnetic and intralayer AFM coupling, similar as for other cases.}
	\label{fig:order}
\end{figure}

\begin{table}[ht]
	\setlength{\tabcolsep}{0.02\textwidth}
	\begin{tabular}{|c|c|c|}
		\hline \text {{\bf Stacking orders}} & \text { A-A stacking } & \text { A-B stacking } \\
		\hline \text { FM-FM } & \text { FM} & \text { FM } \\
		\hline \text { FM-AFM } & \text { AFM } & \text { AFM } \\
		\hline \text { AFM-FM } & \text{ CSML } & \text { AFM } \\
		\hline \text { AFM-AFM } & \text { AFM } & \text{ CSML } \\
		\hline
	\end{tabular}
	\caption{{\bf Spin-splitting for different stacking orders.} FM, AFM and CSML refer to the ferromagnetism, spin-degenerate antiferromagnetism and spin-splitting CSML systems.}
	\label{splitting}
\end{table}

\begin{figure}[ht]
	\centering
	\includegraphics[width=0.9\linewidth]{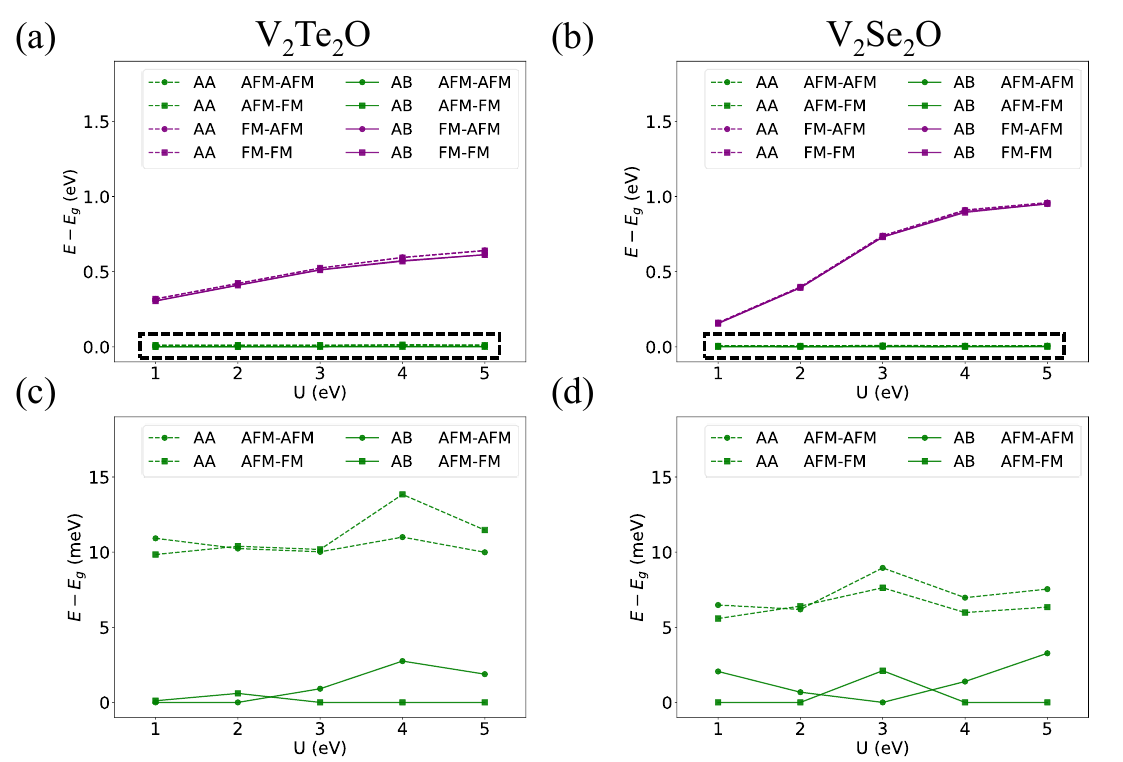}
	\caption{{\bf Energy difference of different stacking orders compared with the ground states.}}
	\label{fig:energy}
\end{figure}

\begin{table*}[htbp]
	\setlength{\tabcolsep}{0.015\textwidth}
	\begin{tabular}{|c|c|c|c|c|c|c|c|c|}
		\hline 
		Stacking orders & \multicolumn{4}{c|}{ A-A stacking } & \multicolumn{4}{c|}{ A-B stacking } \\
		\hline 
		Magnetic orders & AFM-AFM & AFM-FM & FM-AFM & FM-FM & AFM-AFM & AFM-FM & FM-AFM & FM-FM \\
		\hline $\mathrm{U}=1$ & 10.920& 9.840 &318.160 &318.950 &0.000 &0.110 &305.920 &304.380 \\
		\hline $\mathrm{U}=2$ &10.240& 10.390& 419.700& 422.660& 0.000& 0.600& 410.930& 408.870 \\
		\hline $\mathrm{U}=3$ &10.020& 10.180& 523.870& 523.880& 0.910& 0.000& 513.140& 510.990\\
		\hline $\mathrm{U}=4$ &11.000& 13.840& 595.090& 593.610& 2.750& 0.000& 573.300& 569.210\\
		\hline $\mathrm{U}=5$ &9.990& 11.470& 641.580& 639.330& 1.880& 0.000& 612.330& 612.840\\
		\hline
	\end{tabular}
	\caption{Total energy difference $\Delta E$ (meV), taking with respect to the energy of ground state, for different stacking orders, magnetic orders and at different U values of V$_2$Te$_2$O. }
	\label{energy_VTO}
\end{table*}

\begin{table*}[htbp]
	\setlength{\tabcolsep}{0.015\textwidth}
	\begin{tabular}{|c|c|c|c|c|c|c|c|c|}
		\hline 
		Stacking orders & \multicolumn{4}{c|}{ A-A stacking } & \multicolumn{4}{c|}{ A-B stacking } \\
		\hline 
		Magnetic orders & AFM-AFM & AFM-FM & FM-AFM & FM-FM & AFM-AFM & AFM-FM & FM-AFM & FM-FM \\      
		\hline $\mathrm{U}=1$ & 6.480& 5.580& 160.160& 159.850& 2.060& 0.000& 157.170& 154.290 \\
		\hline $\mathrm{U}=2$ & 6.190& 6.410& 397.940& 398.870& 0.680& 0.000& 393.960& 392.220 \\
		\hline $\mathrm{U}=3$ & 8.950& 7.630& 738.950& 737.970& 0.000& 2.110& 730.800& 730.330\\
		\hline $\mathrm{U}=4$ & 6.970& 5.980& 910.830& 907.800& 1.390& 0.000& 900.090& 893.890\\
		\hline $\mathrm{U}=5$ & 7.540& 6.340& 960.070& 955.320& 3.270& 0.000& 950.390& 951.580\\
		\hline
	\end{tabular}
	\caption{Total energy difference $\Delta E$ (meV), taking with respect to the energy of ground state, for different stacking orders, magnetic orders and at different U values of V$_2$Se$_2$O. }
	\label{energy_VSO}
\end{table*}

\subsection{TB model of bilayer systems}\label{appendix_bilayer}

The bilayer Hamiltonian is expressed as:

\begin{equation*}
	H_{B} = \begin{pmatrix}
		H_0 + \alpha E & H_{c} \\
		H_{c}^{\dagger}  & H_0 - \alpha E
	\end{pmatrix},
\end{equation*}

\begin{equation}
	H_{c} = \begin{pmatrix}
		t_{c}^{1} & 0 & 0 & 0 \\
		0 & t_{c}^{2} & 0 & 0 \\
		0 & 0 & t_{c}^{1} & 0  \\
		0 & 0 & 0 & t_{c}^{2}
	\end{pmatrix},
\end{equation}
where $H_0$ is the four-band models without SOC, $\alpha E$ as the potential term coupling to electric field. We set $t_{c}^{1}=0.002$, $t_{c}^{2}=0.001$ and $\alpha E = 0.1$ eV for qualitative calculations in the main text.

\subsection{Full expression of strain-dependent terms}\label{appendix_strain}
In the absence of strain, the two sublattices in V\(_2\)X\(_2\)O are connected by the crystal symmetry \(\mathcal{C}\) (e.g., the mirror operation \(\mathcal{M}_{\bar{x}y}\)), which enforces equivalence between symmetry-related hopping terms in real space:
\[
\begin{aligned}
t_{xz}^{1x} 
&= \langle d_{xz}^{A,\uparrow}(\mathbf{r}) \,|\, \hat{H} \,|\, d_{xz}^{A,\uparrow}(\mathbf{r} + a\hat{x}) \rangle \\[4pt]
&= \langle d_{xz}^{A,\uparrow}(\mathbf{r}) \,|\, \mathcal{M}_{\bar{x}y}^{-1} \hat{H} \mathcal{M}_{\bar{x}y} \,|\, d_{xz}^{A,\uparrow}(\mathbf{r} + a\hat{x}) \rangle \\[4pt]
&= \langle d_{yz}^{B,\downarrow}(\mathbf{r}) \,|\, \hat{H} \,|\, d_{yz}^{B,\downarrow}(\mathbf{r} + b\hat{y}) \rangle 
= t_{yz}^{1y}.
\end{aligned}
\]

When an external strain breaks the pairing symmetry \(\mathcal{M}_{\bar{x}y}\), however, the equivalence between \(t_{xz}^{1x}\) and \(t_{yz}^{1y}\) is lifted. As noted in the main text, strain can induce two distinct effects: a uniform change that preserves the original symmetry, and an alternating distortion that breaks it explicitly. To systematically distinguish these contributions, we decompose the strain-induced parameter changes into symmetric and antisymmetric combinations:
\begin{align*}
	t_{ij}^0(\epsilon) &= [t_{ij}^1(\epsilon) + t_{ij}^2(\epsilon)]/2 = t_{ij}^0(0) + \frac{\partial t_{ij}^0}{\partial \epsilon_{\alpha \beta}}\epsilon_{\alpha \beta} \\
	& = t_{ij}(0) + \Lambda_{\alpha \beta}^0\epsilon_{\alpha \beta}\\
	t_{ij}^{\prime}(\epsilon) &= [t_{ij}^1(\epsilon) - t_{ij}^2(\epsilon)]/2 =  \frac{\partial t_{ij}^{\prime}}{\partial \epsilon_{\alpha \beta}}\epsilon_{\alpha \beta}  \\
	& = \Lambda_{\alpha \beta}^{\prime}\epsilon_{\alpha \beta}
\end{align*}
where \(t^1\) and \(t^2\) denote the symmetry-paired equivalent hopping terms without strain, and \(\Lambda_{\alpha \beta}^0\) and \(\Lambda_{\alpha \beta}^{\prime}\) correspond to the uniform change in the band dispersion and the symmetry-breaking term, respectively.

Hence, the strained Hamiltonian of V$_2$X$_2$O family is expressed as:
\begin{equation*}
	H (\epsilon)= H_0 + H^{\Lambda^0} \cdot \epsilon + H^{\Lambda^{\prime}} \cdot \epsilon,
\end{equation*}
where the full expressions of $H^{\Lambda}$ ($\Lambda=\Lambda^0 $ or $ \Lambda^{\prime}$) are:
\begin{equation}
	H^{\Lambda}= \begin{pmatrix}
		H_{xz, \uparrow}^{\Lambda} & 0 & 0 & 0 \\
		0 & H_{xy, \uparrow}^{\Lambda} & 0 & 0 \\
		0 & 0 & H_{yz, \downarrow}^{\Lambda} & 0 \\
		0 & 0 & 0 & H_{xy, \downarrow}^{\Lambda}
	\end{pmatrix},
\end{equation}

\begin{table}[htbp]
	\setlength{\tabcolsep}{0.009\textwidth}
	\begin{tabular}{l|cccccc}
		\hline \hline
		&$e_{xz}$ & $t_{xz}^{1x}$ & $t_{xz}^{1y}$ & $t_{xz}^{2}$ & $t_{xz}^{3x}$& $t_{xz}^{3y}$\\
		&$e_{xy}$ & $t_{xy}^{1x}$ & $t_{xy}^{1y}$ &$ t_{xy}^{2}$ & $t_{xy}^{3x}$& $t_{xy}^{3y}$\\
		\hline
		$\Lambda_{xx}^0$&0.300&0.554&-0.216&-0.106&-0.108&0.023\\
		&0.576&-&-0.321&0.237&-0.202&-\\
		\hline
		$\Lambda_{xx}^{\prime}$&-1.843&1.098&0.281&0.053&0.029&-0.027\\ 
		&-1.418&0.906&0.312&-&-0.609&0.217\\
		\hline
		$\Lambda_{yy}^0$&0.300&	0.554&	-0.216&	-0.106&	-0.108&0.023\\
		&0.576	&-&	-0.321&	0.237&	-0.202&	-\\
		\hline
		$\Lambda_{yy}^{\prime}$&1.843&	-1.099	&-0.281&-0.054&-0.029	&0.0271\\ 
		&1.418	&-0.907&	-0.313	&-	&0.609	&-0.216\\
		\hline \hline
	\end{tabular}
	\caption{Fitted strain-dependent coefficients for Rb intercalated V$_2$Te$_2$O}
	\label{parameter_strain}
\end{table}

with the matrix elements of  $H^{\Lambda^0}$ are given by:
\begin{align*}
	H_{xz \uparrow}^{\Lambda^0} & = \Lambda_{\alpha \beta}^{0}(e_{xz}) + 4 \Lambda_{\alpha \beta}^{0}(t_{xz}^{2})\cos k_x\cos k_y \\&+ 2 \Lambda_{\alpha \beta}^{0}(t_{xz}^{1x}) \cos k_x+2 \Lambda_{\alpha \beta}^{0}(t_{xz}^{1y})\cos k_y \\&+ 2 \Lambda_{\alpha \beta}^{0}(t_{xz}^{3x}) \cos k_x+2 \Lambda_{\alpha \beta}^{0}(t_{xz}^{3y})\cos k_y, \\
	H_{xy \uparrow}^{\Lambda^0} & =  \Lambda_{\alpha \beta}^{0}(e_{xy})+4 \Lambda_{\alpha \beta}^{0}(t_{xy}^{2}) \cos k_x \cos k_y \\& + 2\Lambda_{\alpha \beta}^{0}(t_{xy}^{1x}) \cos k_x+2\Lambda_{\alpha \beta}^{0}(t_{xy}^{1y}) \cos k_y\\&+ 2\Lambda_{\alpha \beta}^{0}(t_{xy}^{3x}) \cos k_x+2\Lambda_{\alpha \beta}^{0}(t_{xy}^{3y}) \cos k_y ,\\
	H_{y z \downarrow}^{\Lambda^0}&=\Lambda_{\alpha \beta}^{0}(e_{xz})+ 4 \Lambda_{\alpha \beta}^{0}(t_{xz}^{2}) \cos k_x \cos k_y\\&+ 2 \Lambda_{\alpha \beta}^{0}(t_{xz}^{1x}) \cos k_y+2 \Lambda_{\alpha \beta}^{0}(t_{xz}^{1y}) \cos k_x \\& + 2 \Lambda_{\alpha \beta}^{0}(t_{xz}^{3x}) \cos k_y+2 \Lambda_{\alpha \beta}^{0}(t_{xz}^{3y}) \cos k_x,  \\
	H_{x y \downarrow}^{\Lambda^0}&=\Lambda_{\alpha \beta}^{0}(e_{xy}) +4 \Lambda_{\alpha \beta}^{0}(t_{xy}^{2}) \cos k_x \cos k_y \\& + 2 \Lambda_{\alpha \beta}^{0}(t_{xy}^{1x}) \cos k_y+2 \Lambda_{\alpha \beta}^{0}(t_{xy}^{1y}) \cos k_x\\& + 2 \Lambda_{\alpha \beta}^{0}(t_{xy}^{3x}) \cos k_y+2 \Lambda_{\alpha \beta}^{0}(t_{xy}^{3y}) \cos k_x, \\
\end{align*}
and the matrix elements of  $H^{\Lambda^{\prime}}$ are given by:
\begin{align*}
	H_{xz \uparrow} ^{\Lambda^{\prime}}& = \Lambda_{\alpha \beta}^{\prime}(e_{xz})+4 \Lambda_{\alpha \beta}^{\prime}(t_{xz}^{2})\cos k_x\cos k_y\\&+ 2 \Lambda_{\alpha \beta}^{\prime}(t_{xz}^{1x}) \cos k_x+2 \Lambda_{\alpha \beta}^{0}(t_{xz}^{1y})\cos k_y\\& + 2 \Lambda_{\alpha \beta}^{\prime}(t_{xz}^{3x}) \cos k_x+2 \Lambda_{\alpha \beta}^{0}(t_{xz}^{3y})\cos k_y, \\
	H_{xy \uparrow} ^{\Lambda^{\prime}}& =  \Lambda_{\alpha \beta}^{\prime}(e_{xy})+4 \Lambda_{\alpha \beta}^{\prime}(t_{xy}^{2}) \cos k_x \cos k_y\\&+ 2\Lambda_{\alpha \beta}^{\prime}(t_{xy}^{1x}) \cos k_x+2\Lambda_{\alpha \beta}^{\prime}(t_{xy}^{1y}) \cos k_y\\&+ 2\Lambda_{\alpha \beta}^{\prime}(t_{xy}^{3x}) \cos k_x+2\Lambda_{\alpha \beta}^{\prime}(t_{xy}^{3y}) \cos k_y, \\
	H_{y z \downarrow}^{\Lambda^{\prime}}&=-\Lambda_{\alpha \beta}^{\prime}(e_{xz})-4 \Lambda_{\alpha \beta}^{\prime}(t_{xz}^{2}) \cos k_x \cos k_y\\&- 2 \Lambda_{\alpha \beta}^{\prime}(t_{xz}^{1x}) \cos k_y-2 \Lambda_{\alpha \beta}^{\prime}(t_{xz}^{1y}) \cos k_x\\&- 2 \Lambda_{\alpha \beta}^{\prime}(t_{xz}^{3x}) \cos k_y-2 \Lambda_{\alpha \beta}^{\prime}(t_{xz}^{3y}) \cos k_x, \\
	H_{x y \downarrow}^{\Lambda^{\prime}}&=-\Lambda_{\alpha \beta}^{\prime}(e_{xy})-4 \Lambda_{\alpha \beta}^{\prime}(t_{xy}^{2}) \cos k_x \cos k_y\\&- 2 \Lambda_{\alpha \beta}^{\prime}(t_{xy}^{1x}) \cos k_y-2 \Lambda_{\alpha \beta}^{\prime}(t_{xy}^{1y}) \cos k_x\\&- 2 \Lambda_{\alpha \beta}^{\prime}(t_{xy}^{3x}) \cos k_y-2 \Lambda_{\alpha \beta}^{\prime}(t_{xy}^{3y}) \cos k_x.
\end{align*}

The fitted values of $\Lambda_{\alpha \beta}^{0}(t)$ and $\Lambda_{\alpha \beta}^{\prime}(t)$ for $\epsilon_{xx}$ and $\epsilon_{yy}$ are listed in the Table. \ref{parameter_strain}.

\subsection{Methods}
\begin{table}[bp]
	\begin{tabular}{|c|c|c|c|}	
		\hline 
		Material & $a=b$ $(\AA)$  & Material & $a=b$ $(\AA)$ \\
		\hline 
		$\mathrm{RbV}_2\mathrm{Te}_2\mathrm{O}$ & 4.020& $\mathrm{RbV}_2\mathrm{Se}_2\mathrm{O}$ & 3.942  \\
		\hline 
		$\mathrm{KV}_2\mathrm{Te}_2\mathrm{O}$ & 4.008 & $\mathrm{KV}_2\mathrm{Se}_2\mathrm{O}$ & 3.952 \\
		\hline 
		$\mathrm{V}_2\mathrm{Te}_2\mathrm{O}$ & 3.928 & $\mathrm{V}_2\mathrm{Se}_2\mathrm{O}$ & 3.887 \\
		\hline 
	\end{tabular}
	\caption{The lattice constant of V$_2$X$_2$O family used in calculations.}
\end{table}

The first-principles calculations of the electronic structure were performed using the density functional theory framework as implemented in the Vienna {\em ab initio} simulation package~\cite{kresse1996efficiency, kresse1996efficient}. The projector-augmented wave potential was adopted with the plane-wave energy cutoff set to 440\,eV (convergence criteria $10^{-6}\,$eV). The exchange-correlation functional of the Perdew–Burke–Ernzerhof type was used~\cite{perdew1996generalized} with a 11$\times$11$\times$9 gamma-centered Monkhorst–Pack mesh. We constructed the Hamiltonian of a Wannier tight-binding model in Berry curvature calculation using the {\em WANNIER90} interface~\cite{mostofi2008wannier90}, including the V $d$ and Se/Te $p$ orbitals.

\bibliography{bibliography}
\end{document}